\documentclass[aps,prb,preprint,superscriptaddress,floatfix]{revtex4}
\usepackage{amsmath,amsthm}
\usepackage{graphicx,layout}
\newcommand{\feff}{{\footnotesize FEFF}}

\newcommand{\bra}[1]{\left < #1 \right |}
\newcommand{\ket}[1]{\left | #1 \right >}

\newcommand{\matrixel}[3]{\left < #1 \, \vrule #2 \, \vrule \,  #3 \right >}
\newcommand{\vecr}{\vec r}
\newcommand{\vecrp}{\vec r\,'}

\newcommand{\alox}{${\rm Al}_2{\rm O}_3$}
\newcommand{\colvec}[2]{\left (  \begin{array}{c} #1 \\ #2 \end{array} \right )}

\graphicspath{{./figs/}} 
\date{\today}

\usepackage{graphics}

\newcommand{\beq}{\begin{equation}}
\newcommand{\eeq}{\end{equation}}
\newcommand{\beqa}{\begin{eqnarray}}
\newcommand{\eeqa}{\end{eqnarray}}

\begin{document}

\title{Density matrix calculation of optical constants from optical to x-ray
frequencies}

 \author{M. P. Prange}
 \affiliation{Department of Physics, University of Washington, Seattle, WA 98195}

 \author{J. J. Rehr}
 \affiliation{Department of Physics, University of Washington, Seattle, WA 98195}
 
 \author{G. Rivas}
 \affiliation{Instituto de Ingenier\'ia y Tecnolog\'ia,
Universidad Aut\'onoma de Ciudad Ju\'arez, Ju\'arez, 32310 Mexico}
 
 \author{J. J. Kas} 
 \affiliation{Department of Physics, University of Washington, Seattle, WA 98195}

 \author{John W. Lawson} 
 \affiliation{Mail Stop 229-1, NASA Ames Research Center, Moffett Field, California 94035}
\begin{abstract}
We present a theory of linear optical constants based on a single-particle
density matrix and implemented in an extension of
the real-space multiple scattering code \feff.
This approach
avoids the need to compute wave-functions explicitly, and yields efficient
calculations for frequencies ranging from the IR to hard x-rays,
and applicable to arbitrary aperiodic systems.
Our approach is illustrated with calculations of optical properties
and applications
for several materials.
\end{abstract}

\pacs {PACS numbers}

\maketitle

\begin{section}{Introduction}
This work is primarily concerned with theoretical calculations of
{\it optical constants}, which are obtained from the long-wavelength limit
$\vec q \rightarrow 0$ of the dielectric function
$\epsilon(\vec q, \omega)$.
These include the complex dielectric
constant $\epsilon(\omega)$, the complex index of refraction,
the energy-loss
function, the photoabsorption coefficient, the photon scattering amplitude
per atom, and the optical reflectivity.
The {\it ab initio} calculation of these optical properties for
arbitrary materials has been a long-standing problem in
condensed-matter physics. \cite{tddftvmbpt,nozieres,ehrenreich,adler,wiser}
Thus in practice, these properties are often estimated from atomic
calculations or taken from tabulated sources.\cite{hoc,elam, henke, yeh}
However, such tabulations are available only
for a small number of materials over limited spectral ranges.
Thus we aim to develop an efficient and widely applicable method covering a
broad range of frequencies, thereby providing a practical
alternative or complement to tabulated data.

The theory of dielectric response of crystalline systems has been developed
extensively over the past several decades, \cite{tddftvmbpt}  following
pioneering works of Nozi\`eres and Pines,~\cite{nozieres} Ehrenreich and
Cohen,~\cite{ehrenreich} Adler,~\cite{adler} and Wiser.~\cite{wiser} These
works developed the self-consistent field approach for the dielectric
function within the time-dependent Hartree approximation, also known as the
random phase approximation (RPA).
Subsequently the theory has been extended  to include exchange effects
within the time-dependent density functional theory
(TDDFT).\cite{zangsov,rungegross}  
While ground-state DFT calculations are now routine, theoretical methods for
accurate calculations of optical spectra are still not widely available.
More elaborate theories have been developed that take into account
quasi-particle effects and particle-hole interactions based on the
Bethe-Salpeter 
equation (BSE),\cite{shamkohn,strinati,rohlfing}  but these are even
more computationally demanding.

In order to remedy this situation we have developed an efficient,
real-space approach within the
adiabatic local-density approximation that can be applied to arbitrary
condensed systems over a broad range of frequencies from the visible
to hard x-rays. Our approach is based on a density matrix 
formulation within an effective single-particle theory.
This approach is a generalization of the real-space Green's function
method implemented in the \feff\ codes
that includes both core- and valence-level spectra.
Our work is intended to
extend the capabilities and ease-of-use of \feff\ to enable {\it full spectrum}
output for general aperiodic systems with a quality roughly comparable to
that in currently available tabulated data.~\cite{hoc,elam,henke,yeh} This
effort was begun by one of us 
using an atomic approximation for initial states.~\cite{gildardo} This
approximation is often adequate at high
frequencies, but is unsatisfactory for optical and UV spectra. 

 The remainder of this paper is arranged as follows. Sec.\ II.\ describes
the theoretical formalism behind our approach; Sec.\ III.\ presents typical
results for various optical constants for a number of materials; Sec.\ IV.\
discusses some additional applications and diagnostics, and Sec.\ V.
presents a brief summary and conclusions.

%

\end{section}

 \begin{section} {Theory}
 
\begin{subsection}{Density matrix theory of dielectric response}

We consider the macroscopic linear response of extended systems to an
external electromagnetic field of polarization
$\hat \epsilon$ and frequency $\omega$
\begin{equation}
V_{\rm ext}(t)=
 V_{\rm ext}(\omega) e^{(-i\omega+\delta)t} + cc,
 \label{vext}
\end{equation}
where $\delta$ is a positive infinitesimal corresponding to adiabatic turn-on
of the perturbing potential.  Throughout this work we use Hartree atomic units
($\hbar=m=e^2=a_0=1$) unless otherwise specified.  This perturbation polarizes
the material, inducing a steady-state change $\delta n(\vec
r,\omega)e^{-i\omega t} + cc$ in the microscopic electron density, which leads
to a macroscopic polarization $\vec P(\omega)e^{-i\omega t}+cc$, representing
the average {\it screening} dipole response of the electrons to the applied
field.  For simplicity of discussion, we assume that $\vec P$ has no component
perpendicular to the applied electric field. This is the case for systems of
cubic or higher symmetry in the $q \rightarrow 0$ limit, but relaxing this
restriction poses no computational difficulty. In this case one can define a 
scaler electric susceptibility $\chi$, and the dielectric function is
~\cite{jackson}
\begin{equation}
  \begin{split}
& \epsilon(\omega)  = 1+4 \pi \chi(\omega) \\
& \vec P=\chi\vec E,
  \end{split}
\end{equation}
where $\vec E$ is the electric field.
Our calculations here make use of
an effective single-particle microscopic theory in which the
$N$-electron state of the system at time $t$ is described by a
Slater determinant of time-dependent
single-particle orbitals ${\phi_i(t)}$. Thus the state can
be characterized by the single-particle density matrix $\rho$ which is simply
the projector onto the orbitals:
\begin{equation}
\rho(t)=\sum_{i=1}^{N}\ket{\phi_i(t)}\bra{\phi_i(t)}.
\end{equation}
Their time evolution is governed by the time-dependent Schrodinger equation
\begin{equation}
  i \frac{d}{dt}
  \ket{\phi_i(t)}=H \ket {\phi_i(t)}
  \label{schrodinger eq}
\end{equation}
for the time-dependent Kohn-Sham Hamiltonian
\begin{equation}
H=-\frac{1}{2}\nabla^2+V_{\rm nuc}+V_H+V_{\rm xc}+\Sigma_d
+V_{\rm ext}(t).
\label{hamiltonian}
\end{equation}
The terms in Eq.\ (\ref{hamiltonian}) are respectively the kinetic
energy, the electrostatic attraction to the nuclei $V_{\rm nuc}$,
the Hartree potential $V_H$, the
ground-state exchange-correlation potential $V_{\rm xc}$, the dynamical
contribution to the quasi-particle self-energy correction in the $GW$
plasmon-pole approximation $\Sigma_d$, and the time-dependent external
potential $V_{\rm ext}(t)$ of Equation (\ref{vext}). Here and below, we 
suppress the position dependence of quantities when no confusion will result.  
The time evolution in Eq.\ (\ref{schrodinger eq}) implies the
Liouville equation~\cite{ehrenreich} for the density matrix
\begin{equation}
i\frac{d \rho}{dt}=H\rho-\rho H^\dagger.
\label{liouville}
\end{equation}
In order to obtain the optical constants,
we first linearize this equation with respect to the ground-state
by decomposing the Hamiltonian and density matrix into their values
in the ground-state and parts induced by $V_{\rm ext}$
\begin{equation}
  \begin{split}
    H&=H_{0}+H_{1}(t)= H_{0}+V_{\rm ext}(t)+V_{\rm ind}(t)\\
    \rho&=\rho_{0}+\rho_{1}.
  \end{split}
\end{equation}
$H_1$ consists of the external field and a term $V_{\rm ind}$
due to the response of the electrons.
Second order terms, i.e., 
the products $\rho_{1}H_{1}$ and $H_{1}\rho_{1}$ are discarded.
We assume that the induced potential $V_{\rm ind}$
and hence $H_1$ {\it have the same 
time dependence as} $V_{\rm ext}$.  With these assumptions, 
the time derivative in Eq.\ (\ref{liouville}) becomes
trivial and we can solve 
Eq.\  (\ref{liouville}) for the induced density matrix in
terms of the Kohn-Sham (KS) orbitals $\ket {\phi^{0}_i}$ and
eigenvalues $E_i$ of the ground-state system, 
\begin{equation}
{\rho_{1}(\omega)} 
         =\sum_{i,j}(f_{i}-f_{j})\frac
         {\ket{\phi^{0}_{i}}\matrixel{\phi^{0}_{i}}
         {H_{1}} {\phi^{0}_{j}}\bra{\phi^{0}_{j}}}
         {\omega-(E_{j}-E_{i})+i\delta},
         \label{rho1}
\end{equation}
where $f_{i}=f(E_{i}) \approx \theta(\mu-E_{i})$ is the Fermi occupation
number of state $\ket{\phi_{i}^{0}}$ and $\mu$ is the Fermi level. The
KS orbitals obey the unperturbed Schrodinger equation
\begin{equation}
  i \frac{d}{dt}
  \ket{\phi_i(t)}=H_0 \ket {\phi_i(t)}.
  \label{ks eq}
\end{equation}
The induced electron density $\delta n(\vecr,\omega)$ due to the perturbation $V_{\rm ext}$ is then given by
\begin{equation}
\delta n(\vecr,\omega)=\matrixel{\vecr}{\rho_{1}(\omega)}{\vecr}.
\label{deltan}
\end{equation}
At this point it is convenient to introduce the bare and full susceptibilities
whose local behavior is given by
\begin{equation}
    \delta n(\vecr,\omega)=\matrixel{\vecr }{\chi^{0}
(\omega)H_{1}}{\vecr}=\matrixel{\vecr }{\chi (\omega)V_{\rm ext}}{\vecr}.
\label{chidef}
\end{equation}
Typically, the bare response $\chi^0$ to an external perturbation is first
computed from a single-particle (i.e.  non-interacting) description of the
ground state. The full response $\chi$ of the system
can be related to response $\chi^0$ of some non-interacting reference system.
This procedure gives rise to the
Dyson equation for $\chi$ with an interaction kernel $K$
\begin{equation}
  \chi=\chi^0+\chi^0K\chi =\chi^0(1 - K \chi^0)^{-1}.
\label{chi inverse}
\end{equation}
Methods for computing optical response that start from a single-particle 
description of the ground state can be classified by their approximations to
the particle-hole interaction kernel $K$.
The accuracy of the calculated macroscopic properties reflect 
that of
the non-interacting response and 
the interaction kernel.
Note, in particular, that one needs to
find the frequency-dependent response of the non-interacting system, which 
involves different considerations than those for static,
ground state properties (e.g., the ground state energy and density).

In the crudest approximation $K=0$: the resulting polarizability is that of the
non-interacting reference system and local fields are neglected. In this case
there is no screening, and the single-particle potential is the sum of the
ground-state potential and $V^{\rm ext}$.  An obvious deficiency of the
non-interacting response is that the Coulomb field of the induced
density is neglected. To address this deficiency
Adler~\cite{adler} and Wiser~\cite{wiser} developed formally equivalent theories
of the macroscopic dielectric response of periodic solids based on the
RPA in which $K$ is taken to be the
bare Coulomb interaction.  These theories were originally
built on band structure calculations for periodic materials
in the Hartree approximation, and Hartree local fields were included
through the now termed {\it Adler-Wiser formula}.
In this approach the operator inversion of Eq.\  (\ref{chi inverse}) is
reformulated using the inverse of the microscopic dielectric matrix
$\epsilon_{GG'}(\omega,\vec k)$,
which is then spatially averaged to give the macroscopic response
$\epsilon(\omega)=\lim_{k \to 0} 1/{[\epsilon^{-1}(\vec k, \omega)]_{0,0}}$.
However, the Adler-Wiser dielectric function is that of the
Hartree system and
has the deficiency that the underlying electronic wave function is not
anti-symmetric under particle interchange.

Going beyond RPA thus requires additional exchange-correlation effects in $K$.
There have been efforts along these lines of two types: those based on
time-dependent formulations of density-functional theory (TDDFT), and those
based on many body perturbation theory and the BSE. These 
approaches have been critically compared by
Onida et al.~\cite{tddftvmbpt} By considering excited states from a
quasi-particle viewpoint,~\cite{rohlfing} the interaction kernel can be
decomposed into a direct term $K^D$ which is the Coulomb interaction between
the quasi-particles and an exchange interaction $K^X$,
\begin{equation}
  K=K^X+K^D.
\end{equation}
Expanding Eq.\  (\ref{chi inverse}) in singly-excited (one electron, one
hole) states and taking $K^D$ to be the Coulomb interaction screened by an
effective (microscopic) dielectric function, yields a set of approximations
referred to as the Bethe-Salpeter Equation (BSE). Various screening
models are used ranging from parametrized models (e.g.\ the Levine-Louie
dielectric function) to independent-particle approximations such as the static
RPA. BSE schemes can become computationally demanding since the inverse in Eq.\
(\ref{chi inverse})  must be dealt with in a product basis which can be large.
The differences between the independent-particle excitation energies and
optical spectra and their interacting counterparts are referred to as excitonic
effects. However, the non-locality of the exchange-correlation terms
can be avoided by 
including exchange-correlation effects in $K$ in terms of a
density-functional $f_{\rm xc}$. Then  the approach reduces to
the TDDFT ~\cite{zangsov} where
\begin{equation}
  K(\omega)=v+ f_{\rm xc}(\omega), \quad 
  f_{\rm xc}(\omega)=\frac{\delta V_{\rm xc}}{\delta \rho}.
\end {equation}
Consequently a local approximation to
$v_{\rm xc}$ leads to a local kernel (i.e., $K$ depends only on the diagonal
elements of the real-space single-particle density matrix). This locality
implies that Eq.\ (\ref{chi inverse}) can be expanded in a single-particle
basis, thus circumventing the need for particle-hole states needed for 
the BSE. The
cost of this simplification is that direct information about the particle-hole
interaction (e.g. exciton wave-functions) is only implicit. This makes it
difficult to systematically improve on the local density
approximation (LDA)~\cite{tddftvmbpt}.  Nevertheless, calculations in such
TDLDA frameworks have been carried out for a variety of systems.
\cite{PhysRevLett.88.066404, PhysRevLett.82.1919}
While the TDDFT has achieved good agreement with
experiment for optical spectra in many cases, quantitative agreement at
higher frequencies has been more elusive. Calculations with the BSE
tend to be even more computationally limited.
In addition these methods are built on various ground-state KS
calculations, depending on the system.
 Each approach can work well for a specific class of materials,
but can lose accuracy or applicability for others. Also, the
ground-state methods used were originally developed to calculate static
properties and calculations of frequency-dependent (non-interacting) response
can become cumbersome due to the need for large basis sets and special
exchange-correlation functionals to describe
unoccupied and excited states.

The above difficulties have led us to consider a different approach with 
the goal of developing a general method
for calculations of optical response
that can handle a variety of systems and spectral ranges.
Our approach is based on an extension of real-space multiple scattering
theory (RSMS) in terms of the one-particle density matrix.  The
RSMS approach is well suited to treat arbitrary aperiodic
condensed-matter systems over a very broad
frequency range (from the visible to hard x-rays).
Indeed, this scattering-theoretic approach provides a superior basis
for very high energy spectra where scattering is weak and the approach
converges rapidly.  Further
the approach goes beyond the Born-Oppenheimer approximation and can
include nuclear motion effects in terms of correlated
Debye-Waller factors.~\cite{poiarkova}

In this work, we present calculations within this RSMS approach
using an independent quasi-particle approximation
for the single particle states.
 Comparing Eq.\ (\ref{deltan}) and (\ref{chidef})
gives an expression for the bare response function or susceptibility
\begin{equation}
 \chi^0 (\vec r,\vec r\,',\omega)  
         =\sum_{i,j}(f_{i}-f_{j})\frac
         {\phi^{0}_{i}(\vecr)
         \phi^{0*}_{i}(\vecrp)
         \phi^{0}_{j}(\vecrp)
	 \phi^{0*}_{j}(\vecr) }
         {\omega-(E_{j}-E_{i})+i\delta}.
         \label{chi0 spectral rep}
\end{equation}
Formally the imaginary part of the dielectric function is related to the 
full susceptibility by~\cite{zangsov}
\begin{equation}
\begin{split}
      \epsilon_2 (\omega)&=\frac{4\pi}{V}{\rm  Im} 
      \int{d\vec r\,d\vec r\,'\, 
      {\rm Tr\,}\, d\, \chi (\vec r,\vec r\,',\omega)\, d^{\dagger}}, \\
      \end{split}
  \label{basicresult}
\end{equation}
\noindent
where $V$ is the volume of the system, and
$d=\vec \alpha  \cdot \epsilon_p e^{i\vec k \vec r}$ is the 
transition operator between the incident photon 
of wave vector $\vec k$ and polarization $\epsilon_p$.
In practice the transition operator is replaced by the
truncation to rank-one of its expansion into tensors developed by
Grant,~\cite{grant} which is equivalent to the dipole approximation.

To evaluate Eq.\  (\ref{basicresult}) for both optical
and x-ray frequencies, we must first compute the response function
$\chi^{0}(\vec r,\vec r\,',\omega)$.
Formally Eq.\ (\ref{chi0 spectral rep}) can
be expressed in terms of the single-particle Green's function as
\begin{equation}
\begin{split}
  \chi^0 (\vec r,\vec r\,',\omega) &= 
 \int^{E_F}   \rho(\vec r,\vec r\,',E) 
 G^+(\vec r,\vec r\,',E+\omega) \\
 &+\  \rho(\vec r\,',\vec r,E)G^-(\vec r\,',\vec r,E-\omega)\,dE.
\end{split}
\end{equation}
Using the symmetries $\rho(\vec r,\vec r\,',E)= \rho(\vec r\,',\vec r,E)$ and
$G^-(\vec r,\vec r\,',E)= [G^+(\vec r\,',\vec r,E)]^*$ on the real
$E$-axis we can express the results entirely in terms of the one-particle
density matrices $\rho(E)$ 
\begin{equation}
  -\frac{{\rm  Im}\,\chi^0 }{\pi}=\int^{E_F}_{E_F-\omega}
                    \rho  (\vec r\,',\vec r,E)
                    \rho  (\vec r\,',\vec r,E+\omega)\,dE.
		    \label{chi0}
\end{equation}
In this work we calculate  these density matrices for
energies ranging from the lowest occupied states to very high
energies of order 100 KeV.\cite{review}

\end{subsection}

\begin{subsection}{Multiple scattering Green's function}
Our calculations use an independent electron model in which
each electron moves in an effective quasi-particle scattering
potential $V(\vec r)$
which implicitly includes a dynamic self-energy correction
$\Sigma_d(E)$ to the ground state exchange and correlation potential.  In
this work $\Sigma_d(E)$ is calculated using the local $GW$ plasmon-pole model
of Hedin and Lundqvist.\cite{hl} The potential  $ V(\vec r)=\sum_n v_n(r_n) +
V_{0}$ is taken to be the self-consistent muffin-tin potential for a cluster of
atoms at fixed locations ${\vec R_n}$.  Here $\vec r_n = \vec r-\vec R_n$ is
the position relative to the $n^{\rm th}$ atom, and $V_{0}$ is a constant
interstitial potential. Within  RSMS theory, the 
Green's function for this potential can be written as a
double angular momentum expansion
 \begin{eqnarray}
  G(\vecr,\vecrp,E) &=& -2k \Big [  \sum_{LL'} R_{Ln}(\vecr_n) 
  \tilde G_{Ln,L'n'} 
  \bar R_{L'n'}(\vecrp_{n'}) \nonumber \\ 
  &+&\ \ \  \delta_{n,n'}\sum_L H_{Ln}(\vecr_>) \bar R_{Ln'}(\vecr_<) 
  \Big ], 
 \label{Full G}
 \end{eqnarray}
where  $n$ and $n'$ are the sites nearest $\vecr$ and $\vecrp$ respectively, and
$\vecr_>$ ($\vecr_<$) is the larger (smaller) of the two position vectors.
The terms in equation (\ref{Full G}) 
are the right-hand-side regular and irregular solutions
$R_{Ln}$, $H_{Ln}$
of the spherically symmetric single-site problems and their left-side
counterparts $\bar R_{Ln}$, $\bar H_{Ln}$, the partial-wave phase shifts
$\delta_{ln}$, and the multiple scattering (MS) matrix $G_{Ln,L'n'}$.  
The wave functions are normalized so that in the interstitial region
$R_{Ln}$ coincides with 
$Y_L[ h^+_l e^{i\delta_{ln}} - h^-_l e^{-i\delta_{ln}} ]/2i$, and
$H_{Ln}$ coincides with
$Y_Lh^+_l e^{i\delta_{ln}}$. The bar for the left-sided solutions
indicates that all factors except the Bessel functions are to be complex
conjugated. Eq. (\ref{Full G}) is rederived in the Appendix. As
detailed in the Appendix, all these ingredients except the MS matrix can be
found from the solution of a spherically symmetric single-particle quantum
mechanics problem. The full MS matrix $G$ for the system is found by numerical
matrix inversion (e.g., with the LU or Lanczos algorithms in \feff) with
typical matrix dimensions of order $2\times 10^3$ or using the MS path
expansion.

\subsection{Relativistic basis}
To include relativistic effects such as spin-orbit coupling in our
calculations properly it is necessary to recast the Green's function
in terms of 
spinor solutions to the Dirac equation. In this context it is convenient to 
expand the spin-angular dependence of the one-electron states in the Pauli
spinor-valued spin-orbit eigenfunctions which diagonalize both total and 
orbital angular momentum
   \begin{equation}
     \begin{split}
       &\chi_{K}(\hat r)=\sum_{\sigma=-\frac{1}{2}}^{\frac{1}{2}}
       Y_l^{m_j-\sigma}(\hat r) \phi^{\sigma}
       \left\langle l, \frac{1}{2}, m_j-\sigma, \sigma \vert j, m_j \right\rangle.
     \end{split}
     \label{spin-orbit}
   \end{equation}
 Here $\phi^{\sigma}$ is a Pauli spinor, $K=(\kappa, m_j)$ is a pair of
 relativistic angular momentum quantum numbers, and $\left <l, s, m_l, m_s
 \vert j, m_j \right >$ is a Clebsh-Gordan coefficient. In this work as
 in Ref. [\onlinecite{ankrer}] and [\onlinecite{scatfac}],
we have constructed the scattering matrix
$G_{Ln,L'n'}$ of Eq.\ (\ref{Full G})
using the scattering matrices $t_l$ 
 calculated for the total angular momentum channel $j=l+1/2$.
This matrix is then transformed to the basis of spin-orbit eigenfunctions
using Clebsch-Gordan coefficients. The central-site contribution
Eq.\ (\ref{Gc})
is constructed directly from numerical solutions of the central-site problem
giving a total relativistic Green's function
   \begin{eqnarray}
       G(\vecr,\vecrp,E)&=& -2k \Big [  \sum_{KK'}
       H_{K n}(\vec r_>) \bar R_{K' n'}(\vec r_<)\delta_{KK'}\delta_{nn'}
       \nonumber \\
       &+& R_{K n}(\vecr_n)
       \tilde G_{KnK'n'}
       \bar R_{K' n'}(\vecrp_n) \Big ],
     \label{kappaG}
   \end{eqnarray}
written in terms of right-hand (no bars) and left-hand (bars) solutions of 
the Dirac equation at energy E. These functions are 4-spinors which can be
written in terms of the spin-orbit eigenfunctions:
   \begin{eqnarray}
       &&R_{K n}(\vecr_n) =
       \frac{1}{r_n}\colvec{P_{\kappa}(r_n)\chi_\kappa^{m_j}(\hat r_n)}
       {iQ_{\kappa}(r_n)\chi_{-\kappa}^{m_j}(\hat r_n)} \nonumber \\
       &&\bar R_{K n}(\vecr_n) = 
        \frac{1}{r_n}\colvec
       {P_{\kappa}(r_n)\chi_\kappa^{{m_j}\dagger}(\hat r_n)}
       {-iQ_{\kappa}(r_n)\chi_{-\kappa}^{{m_j}\dagger}(\hat r_n)}^{\rm T},
     \label{RandRbar}
   \end{eqnarray}
where the $T$ in Eq.\ (\ref{RandRbar}) denotes the transposed vector.
The irregular solutions $H$, $\bar H$ take a similar form. These solutions are
normalized by requiring the upper-component radial wave functions to coincide
with $[ h^+_l e^{i\delta_{\kappa n}} - h^-_l e^{-i\delta_{\kappa n}} ]/2i$
(regular solution) or $h^+_l e^{i\delta_{\kappa n}}$ (irregular solution). We
are using the notation of Grant,~\cite{grant} where the reader is referred
for details regarding the numerical solutions $P$, $Q$ appearing in Eq. 
(\ref{RandRbar}). Tamura~\cite{tamura} gives a relevant and illuminating
discussion of solutions to the Dirac equation in spherical coordinates,
although he treats a more general case using different notation.  We have also
transformed to a basis of real spherical harmonics to simplify calculations of
the real-valued density matrices.

\end{subsection}

\begin{subsection}{Complex scattering potential}
The construction of the self-consistent muffin-tin scattering potential
for the one-particle states is described
elsewhere,~\cite{conrad} and we only briefly summarize the process here. First,
a Dirac-Fock solver is used to calculate free-atomic potentials and
densities which are then overlapped to obtain a starting point
for the self-consistency loop. In this loop the one-particle Green's function
for the full multiple scattering problem is calculated, from which
a new electron density is calculated. Finally a new ground state
muffin-tin potential is constructed within the LDA. The loop is iterated to
self-consistency which typically takes about 10-20 iterations. Self-energy
corrections are subsequently added for unoccupied states within the
GW plasmon-pole approximation.
\end{subsection}

\begin{subsection}{Core state response }
At low energies (below the bottom of the valence band), the density matrix
becomes sparse in energy, taking non-zero values only at isolated eigenvalues.
In this regime, it is more computationally efficient to use orbitals to
describe the electronic structure.  Thus, we separate the single particle
density matrix into two energy regions: the core region in which the atomic
approximation is valid and the solid-state region where solid-state corrections
are important,
\begin{equation}
  \rho(E)=
  \begin{cases}
    \rho^{\rm core}(E) & E<E_{\rm cv}, \\
    \rho^{\rm val}(E) & E>E_{\rm cv}. \\
  \end{cases}
\end{equation}
The core-valence separation energy $E_{\rm cv}$ is chosen to be an energy 
away from all KS eigenvalues that separates the two regimes and is set
by default to $-40$ eV, which is typically about 30 eV below the Fermi level.
Above this energy $\rho(E)$ is derived from the single-particle Green's
function as described below. Note that in general there are occupied and
unoccupied states above $E_{\rm cv}$, but there are no unoccupied states below
$E_{\rm cv}$. Similarly, the dielectric function $\epsilon_2(\omega)$can be
separated into contributions $\epsilon_2^{\rm core}(\omega)$ and
$\epsilon_2^{\rm val}(\omega)$ arising from transitions with core and valence
initial states respectively.

The core states are represented by single-particle atomic-like orbitals
$\phi_{\nu}$. Here the index $\nu=(n,i)$ denotes both a site index $n$ and
atomic level index $i$ for the particular bound state at that site
(e.g. 1s, 2s, 2p$_{1/2}$, etc.). We replace $\rho(E)$ in Eq.\
(\ref{chi0}) for $E < E_{\rm cv}$ with 
\begin{eqnarray}
    && \rho_{\rm core}(E)=\sum_{\nu}\rho_{\rm at}^{(\nu)}(E), \\
    && \rho_{\rm at}^{(\nu)}(E) =
    \phi_{\nu}(\vecr)\phi_{\nu}(\vecrp)\delta(E-\epsilon_{\nu}).
    \nonumber
  \label{rhoat}
\end{eqnarray}
Thus we recover an expression equivalent to
Fermi's golden rule for the absorption of light
\begin{eqnarray}
    &&\epsilon_2^{\rm core}(\omega) = \sum_{\nu} \epsilon_2^{\nu}(\omega) \\
&&= \frac{4 \pi }{ \omega} \sum_{\nu}\mathrm{Im}
    \langle\, i\,|\,\hat{d}\,^{\dagger}
    \hat{G}(\omega+\epsilon_i) \hat{d}\,
    |\,i\,\rangle \theta ( \omega+\epsilon_i-\epsilon_F) \nonumber.
    \label{golden rule}
\end{eqnarray}
The initial core states $\ket{\phi_{\nu}}$ and their associated eigenvalues
$\epsilon_{\nu}$ are described accurately by Dirac-Fock atomic states for a
single atomic configuration~\cite{ankzabrer}. 
For energies below $E_{\rm cv} \approx \mu - 30 \ {\rm Ev}\  < V_{0}$ the eigenfunctions of
the central site problem are tightly bound to the central atom; their
wave-functions decay rapidly as a function of the distance from the central
site and can be taken to vanish in all cells except the central cell. This, 
along with the selection rules, limits the elements $G_{KnK'n'}$ (representing
the final states) that contribute to absorption. For core initial states, the
final state energy includes the inverse core-hole lifetime $\Gamma_{\nu}$ which
broadens $\epsilon^{\nu}$. The calculation of the density matrix elements
appearing in equation (\ref{basicresult}) is handled differently depending on
the photoelectron energy $E=\omega+\epsilon_{\nu}$. For low-energy (less than
$\approx$ 50 eV$+V_0$) final states $G$ is calculated by FMS just as in the
calculation of $\epsilon_2^{\rm val}$.  At very high energies we again employ
an atomic model and neglect scattering contributions (i.e.  $G_{KnK'n'}=0$ in
Eq.\ (\ref{kappaG})). At intermediate energies (50 eV $+ V_0 \leq E \leq$ 1000 eV) we
use efficient path filters~\cite{zabin} developed to treat EXAFS to find the
dominant terms in the multiple scattering path expansion and sum these
contributions to obtain the necessary $G_{KnK'n'}$ elements,
\begin{equation}
  G= G^0+G^0TG^0+\cdots .
\end{equation}
The calculation of $\epsilon_2^{\rm core}$ is accomplished by looping over the
edges $\nu$ with eigenvalues below $E_{\rm cv}$. For each edge we calculate 
$\epsilon_2^{\nu}$ via FMS, path-expansion, and the atomic approximation on 
appropriate energy grids.  At this stage, correlated Debye-Waller factors can
be included as in conventional XAS calculations using \feff.
\end{subsection}

\begin{subsection}{Valence response}
 Using the formal relation between the density matrix and the
one-particle Green's function $\rho(E)=(-1/\pi){\rm Im}\, G(E)$
one obtains from Eq.\ (\ref{kappaG})
 \begin{equation}
   \rho^{\rm val}(\vecr,\vecrp,E) = \sum_{K,K'}
    R_{Kn}(\vecr) 
   \hat \rho_{Kn,K'n'} R_{K'n'}(\vecrp_{n'}) ,
   \label{separable rho}
 \end{equation}
which is valid for $\vec r$ in cell $n$ and $\vec r'$ in cell $n'$,
where $R_{Kn}(\vecr) = \chi_K(\hat r_n) R_{\kappa n}(r_n)$.
For real energies, the density matrix can be expressed entirely in terms of the 
regular solutions $R_{Kn}$, and the irregular solutions do not enter.
 Below the Fermi level on the real energy axis, the density matrix is a
rapidly varying function of energy. Away from the real axis, however,
the behavior is much smoother.
To both retain the separable form of Eq.\ (\ref{separable
 rho}) and the smoothness obtained by calculating the Green's function away
 from the real axis, we introduce a small broadening 
 $\Gamma$ 
 and 
 renormalize the regular solutions, so that the central atom density matrix 
 gives the same density of states (DOS) in each Norman sphere as the actual
 broadened density matrix:
 \begin{equation}
 \begin{split}
   &\tilde R_{\kappa n}(r;E)=A_{\kappa n}
 (E,\Gamma)\,{\rm Re}\,\left[R_{\kappa n}(r;E+i\Gamma) \right ], \\
   &\int_0^{r^{\rm Nrm}_n}\left[\tilde R_{\kappa n}(r;E)\right ]^2 r^2 dr=
  {\rm Im}\, 
   \int_0^{r_{\rm Nrm}} r^2 dr  \\
   &\times \int_0^{r^{\rm Nrm}_n} {r'}^2 dr' 
   R_{\kappa n}(r_<;E+i\Gamma)H_{\kappa n}(r_>;E+i\Gamma).
 \end{split}
 \end{equation}
This result is a key simplification in our approach.
Here the Norman radius $r^{\rm Nrm}_{n}$ is defined as
the radius of a neutral sphere centered on the
$n^{\rm th}$ atom in the charge distribution formed by overlapping the charge
distributions of the isolated atoms in their solid-state positions. The
separable representation of the density matrix in Eq.\ (\ref{separable rho})
permits a separation of the double spatial integral in Eq.\ (\ref{basicresult})
into a product of two one-dimensional integrals. 
To complete the spatial integral in
Eq.\  (\ref{basicresult}), we make the approximation that the spherical Norman
cells ${n}$ partition space and write the full integrals as sums of integrals
over individual cells 
\begin{equation}
\int{d\vec r}\, \longrightarrow \sum_{n}\int_{\vec r \in n}{d\vec r}\,
=\sum_{n}\int_{0}^{r_{N}^{(n)}}r_n^2dr_n \int d\Omega_n.
\end{equation}
The dipole matrix elements at each site $n$ are defined as
\begin{equation}
  M^{n}_{K,K\,'}(E,E\,')=\int_{\vec r \in n}{d\vec r}\,
  \tilde R_{Kn}(\vecr; E)\, d\,  \tilde{\bar R}_{Kn} (\vecr;E\,')).
\end{equation}
In the dipole approximation the matrix elements vanish except for transitions
with $j\,'=j\pm 1$. Left (right) circularly polarized light only induces
transitions with $m_j\,'=m_j+1$ ($m_j\,'=m_j-1$). Thus the transition
matrix $M$ is
sparse. Relaxing the dipole approximation is straightforward. Doing so 
introduces additional non-zero elements to $M$.  With these conventions, the
contribution to the spectrum from the response of the valence states (i.e.
those occupied single-particle states with eigenvalues above $E_{\rm cv}$) is
given entirely in terms of density matrices and matrix elements,
  \begin{eqnarray}
      \epsilon_2^{\rm val} (\omega) &&=\frac{4\pi}{V}
      \int^{E_F}_{E_F-\omega} dE \,
      \sum_{n,n'} {\rm Tr \,}
       \rho_{nn'}(E)M_{n'}(E,E+\omega) \nonumber \\
       && \times \rho_{n'n}(E+\omega)M_{n}^{\rm T}(E+\omega,E),
    \label{final result}
  \end{eqnarray}
where $\rho_{nn'}$ and $M_n$ are matrices in a truncated relativistic
angular momentum ${K=(\kappa,\mu)}$-space. 
By symmetry, the sum over sites $n$ in Eq.\  (\ref{final result}) can be
reduced to a sum over inequivalent sites in the solid.
To compute $\epsilon_2^{\rm val}$ we first solve the Dirac equation at each
inequivalent site which yields $T$. Then $G_{Ln L'n'}$ is found by
inverting the full multiple scattering matrix,
and matrix elements $M$ are evaluated using the wave functions
from the calculation of $T$. Finally, Eq.\ (\ref{final result}) is
evaluated using trapezoid rule integration for the energy integrals.

\end{subsection}

\begin{subsection}{Spectrum construction}
With the response of both the valence band and the more tightly bound electrons
calculated, the contribution from each core edge is then interpolated onto a
final output grid and combined with the other core edges and with the valence
contribution:
\begin{equation}
\epsilon_2(\omega)=
\sum_{\nu}\epsilon_2^{(\nu)}(\omega)
+\epsilon_2^{\rm val}(\omega).
\label{eps2}
\end{equation}
\end{subsection}
\end{section}

\begin{section}{Theoretical Optical Constants}

The examples presented here are primarily monatomic crystals (metals
and insulators) with a single inequivalent site. However, the generalization to
heterogeneous materials is straightforward, and an example is also presented for
Al$_2$O$_3$.  Non-periodic materials can be treated by including enough sites
to converge the spectrum. The calculations presented in this section used FMS
matrices truncated at $l=3$ and 147 atoms for all materials except diamond. The
diamond calculation used $l=2$ and 450 atoms. We include Diamond because it is a 
difficult case for our real-space method even though typical $k$-space
calculations of Diamond (such as the plane wave psuedopotential calculation
shown in Fig.~\ref{mu}) use a unit cell containing only two atoms and can be
less computationally demanding. All spectra were obtained by summing the
contributions from 70 atoms.  The response for the valence bands is obtained by
calculating $\rho^{\rm val}$ on a regular energy grid of 200 points. Then the
dipole matrix elements $M(E,E')$ are calculated for all pairs $(E,E')$ with $E$
below the Fermi level and $E'$ above it. Eq.\ (\ref{final result}) is then
evaluated by matrix multiplication and simple numerical integration. To compute
$\epsilon_2^{\rm val}(\omega)$ to high frequencies, we employ an atomic model
of the valence bands based on average band energies and occupations calculated
from $\rho^{\rm val}$. The core state response is first calculated on a set of
five 100 point frequency grids for each core initial state $\kappa$ in the
embedded-atom approximation. The FMS and path-expansion calculations are then
carried out in cluster sizes of around 175 atoms on frequency grids of
approximately 120 points. The contribution to $\epsilon_2$ for each core
initial state and the valence bands are then interpolated onto a large ($5
\times 10^5$ points) frequency grid which spans the full spectrum (e.g.
$10^{-3}$ through $10^6$ eV) and serves as the final output grid. This grid has
a higher density of points at low frequencies and around each core edge.


\begin{subsection}{Dielectric function: Imaginary part}

  The fundamental quantity needed in our calculations of optical response
is the imaginary part of the dielectric function $\epsilon_2(\omega)$ given
by Eq.\ (\ref{eps2}).
 All other optical constants can be obtained in terms of
$\epsilon_2(\omega)$ as described below.
 As illustrative examples our density matrix calculations 
of $\epsilon_2(\omega)$ for Cu
and Au are plotted in 
Fig.~\ref{eps2fig}
compared to
experiment.
\begin{figure}
  \includegraphics{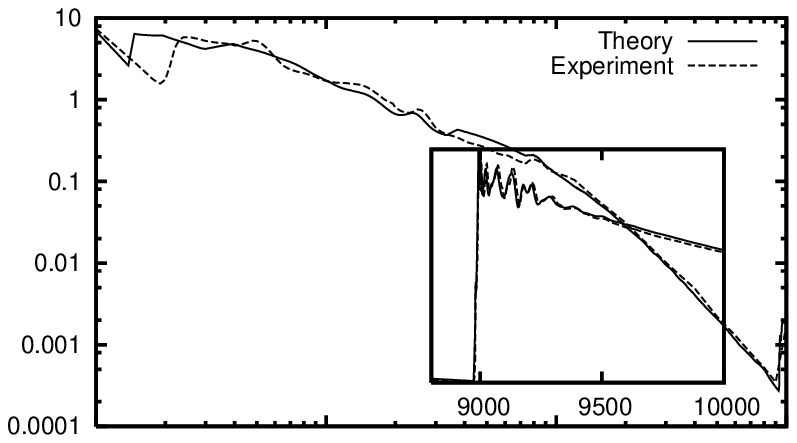} 
  \includegraphics{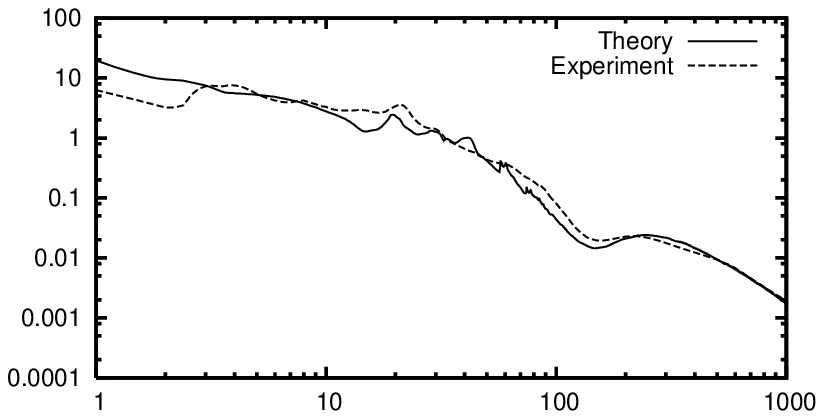} 
  \includegraphics{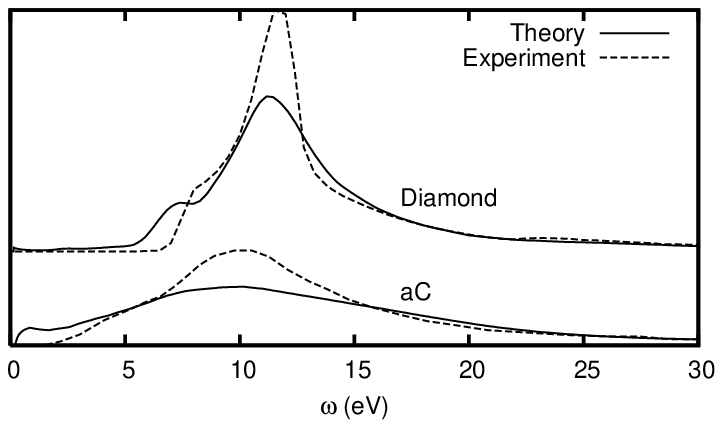} 
  \caption
  {Calculated and experimental $\epsilon_2$ for Cu\cite{desy,datacu} (top), Au\cite{desy} (middle),
  and diamond\cite{hoc} and amorphous C\cite{waidmann} (bottom). In the bottom
  panel the diamond curves have been shifted vertically for clarity.
}
 \label{eps2fig}
\end{figure}
To demonstrate the effects of structural disorder on the dielectric response,
we compare the imaginary part of the dielectric function for Diamond and
amorphous Carbon in Fig.~\ref{eps2}.
Amorphous carbon structures were obtained with a ``melt-and-quench'' algorithm \cite{galli}
using first principles molecular dynamics as implemented in the VASP package.\cite{vasp}
These results, as well as the results presented
below and calculations for other materials, are currently available in both
graphical and tabular form on the \feff ~website.~\cite{web_opcons}

\end{subsection}

\begin{subsection}{Dielectric function: Real part}
 Owing to the analyticity of the dielectric response, 
  the real and imaginary parts of the dielectric function are related by 
  the Kramers-Kronig relation~\cite{jackson}
\begin{equation}
  \epsilon(\omega)=1+\frac{2}{\pi}\, {\cal P}\int_0^{\infty}d\omega\,' 
  \frac{\omega\,' \epsilon_2(\omega\,')}
  {\omega^2-\omega\,'^2}.
  \label{kramerskronig}
\end{equation}
  Here ${\cal P}$ indicates the principal value of the integral. Since the 
  denominator of the integrand in Eq.\ (\ref{kramerskronig}) has a pole
  at $\omega\,'=\omega$ care must be taken when evaluating the transform
  numerically. To evaluate the integral appearing in Eq.\
  (\ref{kramerskronig}) over the interval $(\omega_i,\omega_{i+1})$
  between the $i^{\rm th}$ and $(i+1)^{\rm th}$ grid points 
  we find a linear approximation $\epsilon_2(\omega\,')=m\omega\,'+b$, which allows
  us to rewrite the Kramers-Kronig integral as follows:
 \begin{eqnarray}
   &&{\cal P}\,\int_{\omega_i}^{\omega_{i+1}}d\omega\,' 
  \frac{\omega\,' \epsilon_2(\omega\,')}
  {\omega^2-\omega\,'^2}= 
  m(\omega_{i+1}-\omega_i)+ \\
  &&\frac{b-m\omega}{2}\ln\left(\frac{\omega_{i+1}+
  \omega}{\omega_i+\omega}\right)+
  \frac{b+m\omega}{2}\ln\left(\frac{\omega_{i+1}-\omega}
  {\omega_i-\omega}\right) \nonumber .
 \label{numkk}
\end{eqnarray}
This expression is used to produce $\epsilon_1$ on the same output grid used
for the imaginary part. 
The results of this procedure for diamond, Cu and \alox are plotted 
in Fig.~\ref{eps1metal}.
Even though the numerical
transform Eq.\ (\ref{numkk}) is stable and accurate and (along with the
calculated $\epsilon_2$)
completely determines $\epsilon_1$ via Eq.\ (\ref{kramerskronig}),
we find that
the real part of the dielectric function is more sensitive to errors and
approximations than the imaginary part.

\begin{figure}
 \includegraphics{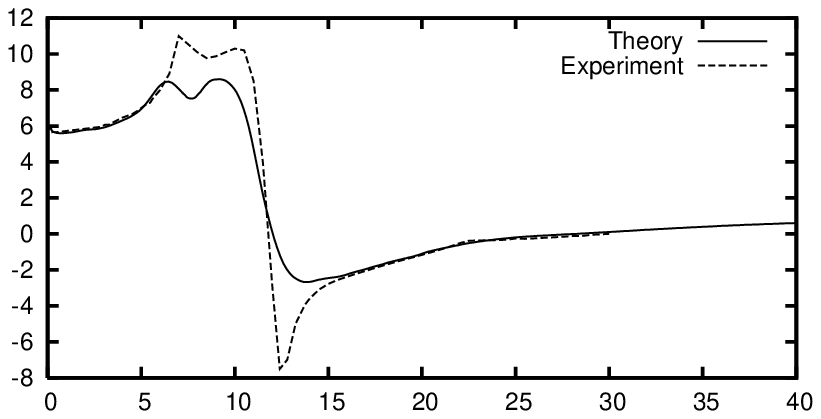}
 \includegraphics{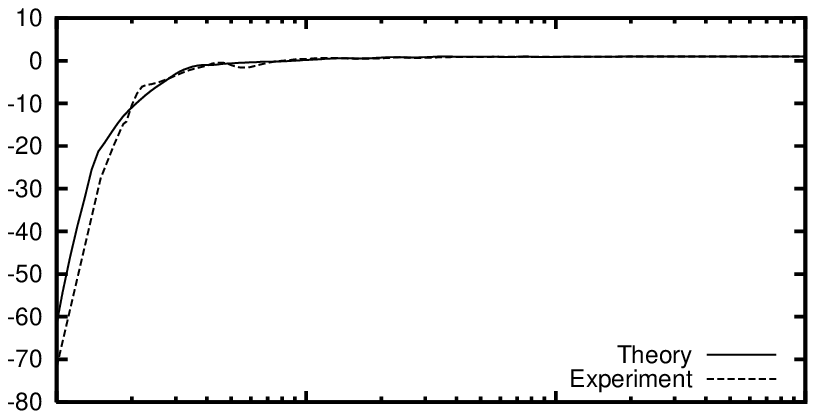}
 \includegraphics{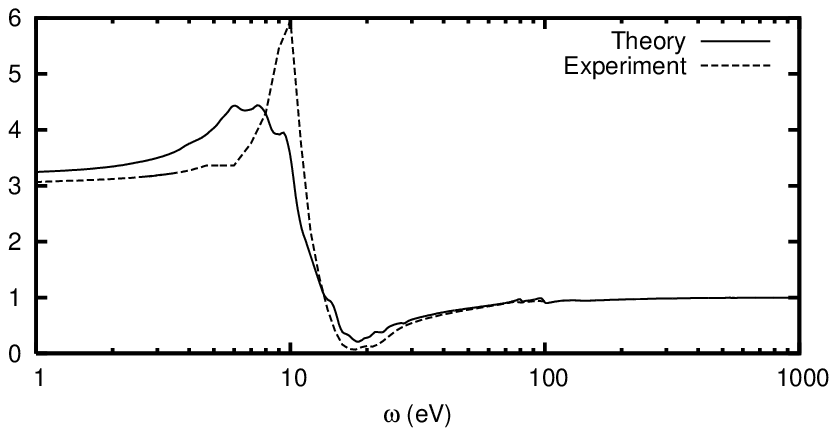}
 \caption
 {Calculated $\epsilon_1$ for diamond (top) and Cu (middle), and \alox (bottom) compared to experiment.~\cite{desy, hoc}} 
 \label{eps1metal}
\end{figure}


\end{subsection}

\begin{subsection}{Energy-loss}
  With both real and imaginary parts of $\epsilon(\omega)$ one can
  easily obtain the energy loss function 
  \begin{equation}
  -{\rm Im}\, \epsilon^{-1}(\omega)=
\frac{\epsilon_2(\omega)}{\epsilon_2^2(\omega)+\epsilon_1^2(\omega)}.
 \label{eloss}
  \end{equation}
This is illustrated for Cu, \alox, and Au in Fig.~\ref{epsinvmetal}.
The loss function is proportional to the
long-wavelength limit of the dynamic structure factor $S(\vec q,\omega)$, which
can be measured by inelastic scattering of either electrons in electron energy
loss spectroscopy (EELS) or photons in non-resonant inelastic x-ray scattering
(NRIXS). Calculations of the latter performed in a framework similar to ours
have recently been reported by Soininen et. al. who only address the response
of core electrons, but at finite $\vec q$.~\cite{soininen} In contrast to
$\epsilon_1$ we find that the loss function is {\it less} sensitive to errors
and approximations in the density matrix than $\epsilon_2$. Onida, et.
al~\cite{tddftvmbpt}., in an illuminating discussion of the differences between
absorption and EELS experiments, have given an explanation of this observation
in terms of the long-range part of the coulomb interaction.
\begin{figure}
 \includegraphics{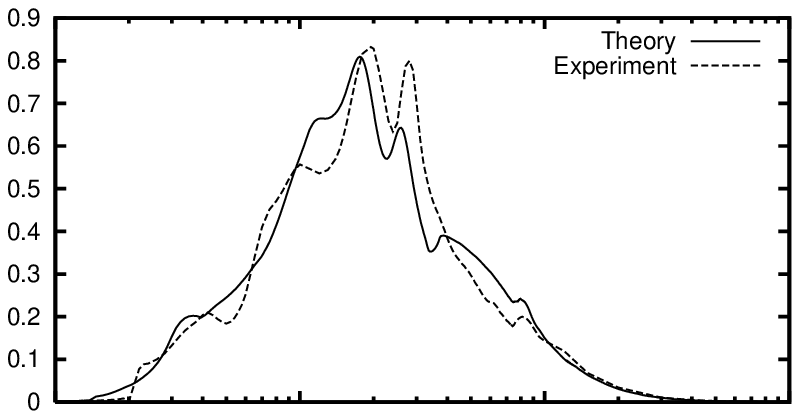}
 \includegraphics{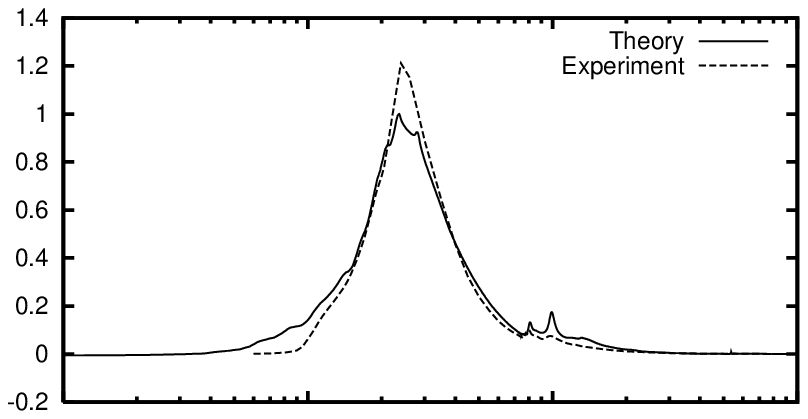}
 \includegraphics{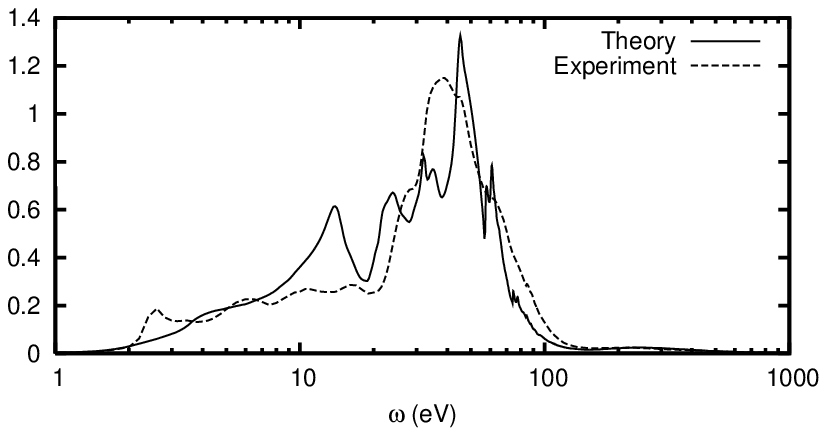}
 \caption
 {Calculated energy-loss function (Eq.\  \ref{eloss}) for Cu (top), \alox (middle), and Au (bottom) compared to experiment.~\cite{desy, hoc}}
 \label{epsinvmetal}
\end{figure}
  \end{subsection}

  \begin{subsection}{Index of refraction}
    The complex index of refraction is simply the square root of the
complex dielectric function
\begin{equation}
n(\omega) + i\kappa(\omega)\equiv\epsilon(\omega)^{1/2} .
\end{equation}
Typical results
for the real part of the index of refraction
are given in Fig.~\ref{n_metal}.

\begin{figure}
 \includegraphics{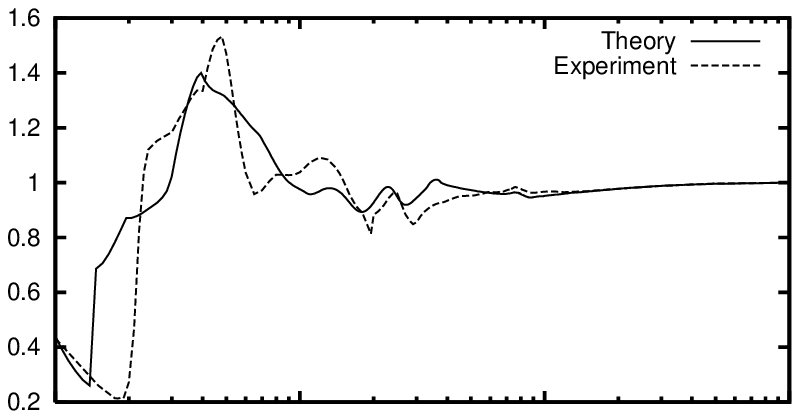}
 \includegraphics{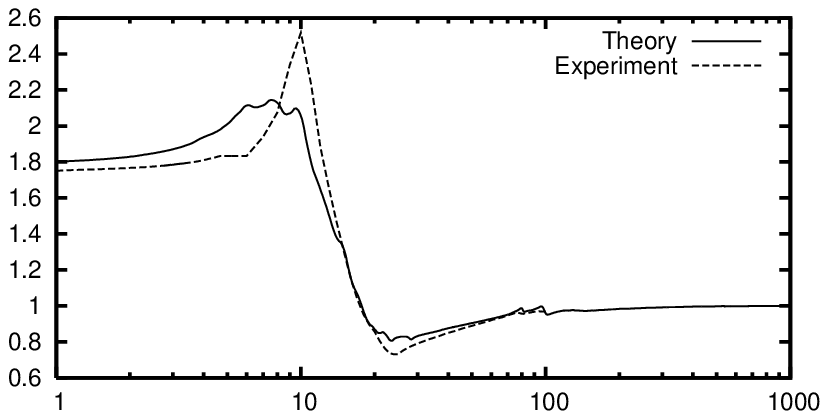}
 \includegraphics{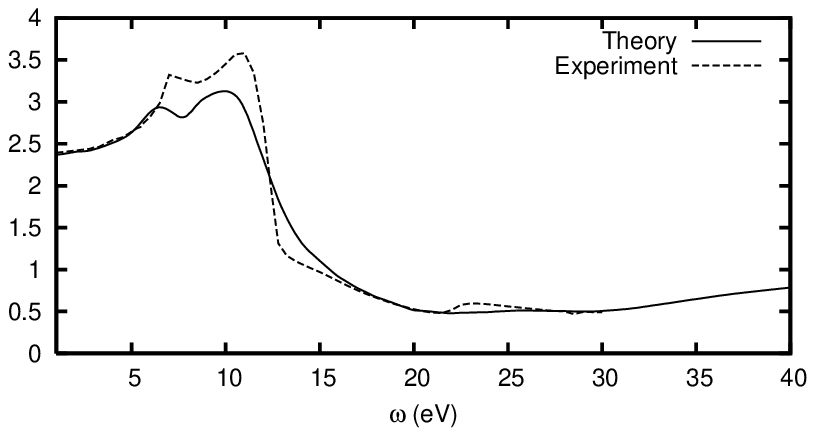}
 \caption
 {Calculated real index of refraction for Cu (top), \alox (middle),  and diamond (bottom) compared to
 experiment.~\cite{desy, hoc}} 
 \label{n_metal}
\end{figure}




  \end{subsection}

  \begin{subsection}{Absorption coefficient}
 The photon absorption coefficient $\mu(\omega)$ is defined as
  the (natural) logarithm of the ratio of the incident and transmitted
intensities for a photon beam across a thin sample, divided by the
 thickness. Theoretically $\mu(\omega)$ can be expressed
in terms of the imaginary part of
the index of refraction $\kappa(\omega)$
\begin{equation}
\mu(\omega) = 2 \frac{\omega}{c}\kappa(\omega).
\end{equation}
Thus, $\mu(\omega)$ is directly measurable with optical absorption
  experiments. Such experiments are currently performed to high
accuracy using synchrotron light sources. We compare our calclulated results
with experiment for several materials and with a calculation based on electronic
structure calculated with ABINIT. This calculation was acomplished using the 
AI2NBSE package developed by Lawler, et. al.~\cite{lawler} which employs a
BSE solver developed at NIST to generate optical spectra. The calculation shown
excludes both local fields and excitonic effects and was generated using a
regular grid of  $8^3$ $k$-points to sample the Brillouin zone, 50 bands, and an
energy cutoff of 30 Hartree for the plane wave basis. The C $1_s$ electrons were
treated with a Troullier-Martins psuedopotential. For a sensible comparision, no
gap corrections were included in either calculation.
 \begin{figure}
  \includegraphics{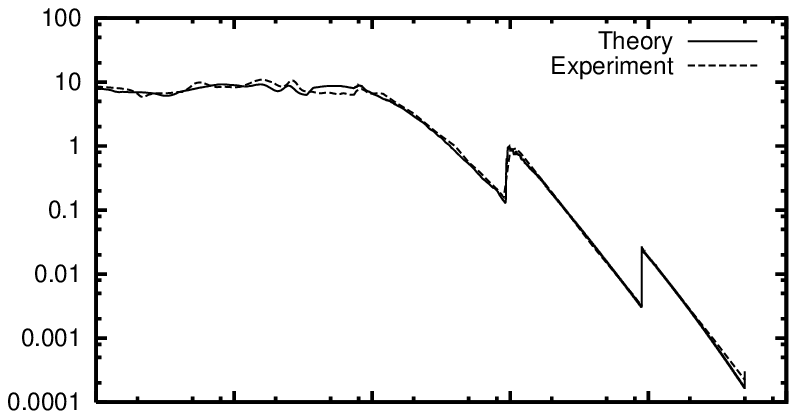}
  \includegraphics{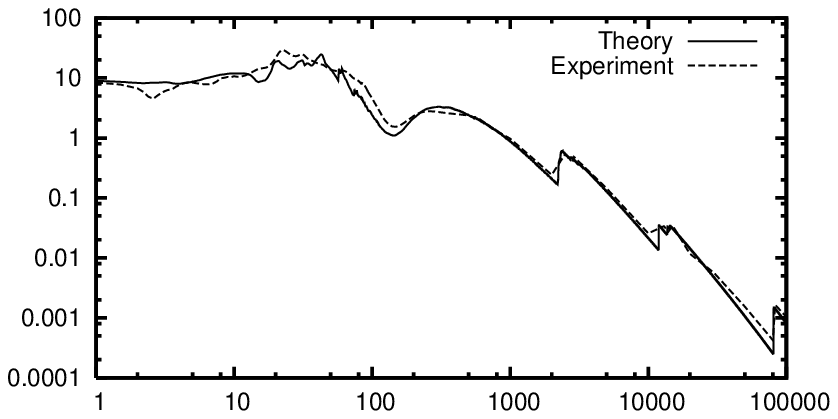}
  \includegraphics{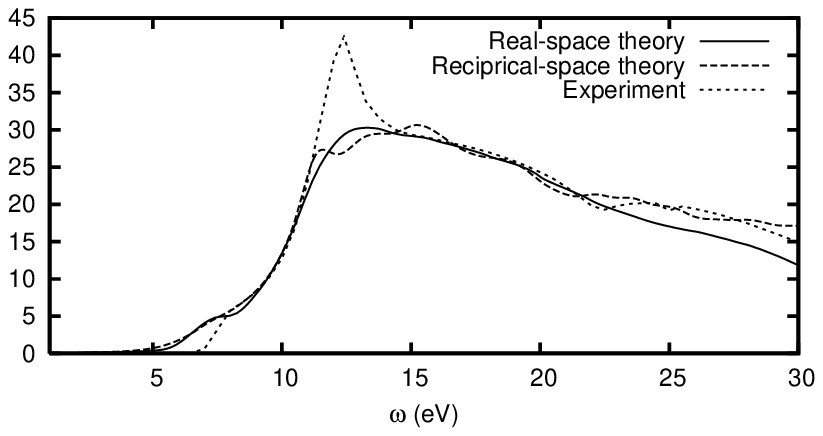}
  \caption
  {Calculated absorption coefficient $\mu$ in inverse cm for Cu (top), Au
  (middle), and diamond (bottom) compared to experiment.~\cite{desy,hoc} The
  calculated diamond result is also compared to a reciprocal-space calculation.}
  \label{mu}
 \end{figure}
\end{subsection}

  \begin{subsection}{Reflectivity}
  An important optical experiment for materials that can be prepared by vapor
  deposition methods is the measurement of the the reflectivity $R$ defined as
  the ratio of the power reflected from a planar face of a sample to the
  incident power. This quantity can be related to the dielectric response of
  the material by considering the boundary conditions satisfied by Maxwell's
  equations at the interface between the sample and vacuum. This procedure
  produces the familiar Fresnel equations~\cite{jackson} relating the
  amplitudes of the transmitted (refracted) and reflected waves to the
  amplitude of the incident wave. As discussed by Stratton,~\cite{stratton} $R$
  can be found by squaring the Fresnel equations. For example, for normal
incidence 
  \begin{equation}
    R(\omega) = \frac{\left [ n(\omega)-1 \right ]^2 +
   \kappa^2(\omega)}{ \left [n (\omega)+1 \right ]^2 + \kappa^2(\omega)}.
  \end{equation}
 The general expression for a lossy material
($\epsilon_2 \neq 0$) and arbitrary
 angle of incidence is complex. However
it is interesting to note that off normal incidence
  $R(\omega)$ has polarization dependence even for isotropic media.

  \end{subsection}

\begin{subsection}{Photon scattering amplitude}
  The Rayleigh forward scattering amplitude $f(\omega)$ for photons
can also be computed from the dielectric function~\cite{scatfact} 
\begin{equation}
  f(\omega)=\frac{\omega}{4\pi r_0 c^2}\frac{V}
  {N}[\epsilon \left ( \omega)-1 \right]. 
\end{equation}
Thus it is straightforward to calculate the x-ray scattering factors 
including anomalous terms using our RSMS approach
$f(\vec q, \omega) = g(q,\omega)+f^{ss}(\vec q,\omega) + f_1(\omega) +
if_2(\omega)$ 
in terms of $f$.
Typical calculations of the real
 and imaginary  parts
of $f(\omega)$ are illustrated in Fig.'s~\ref{fprime} and~\ref{fdblprime_metal}.
 \begin{figure}
 \includegraphics{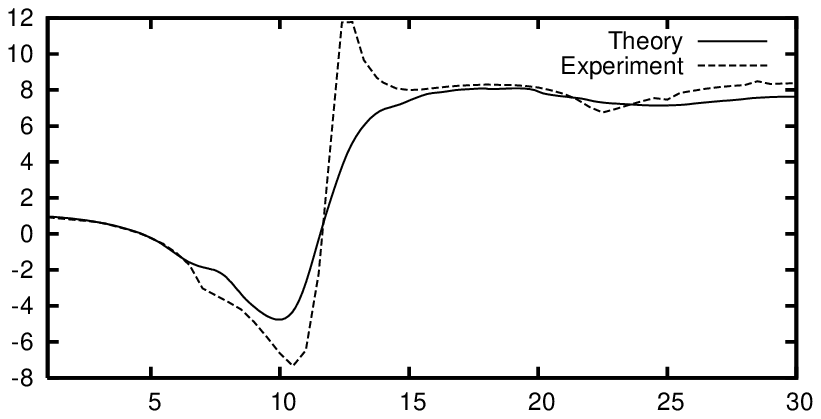}
 \includegraphics{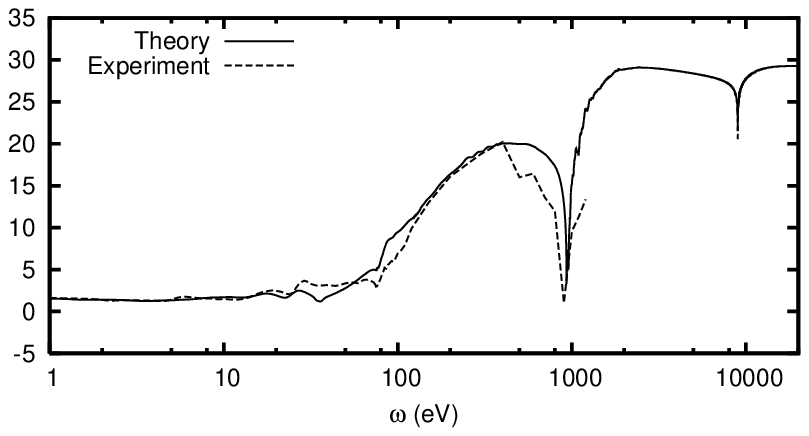}
 \includegraphics{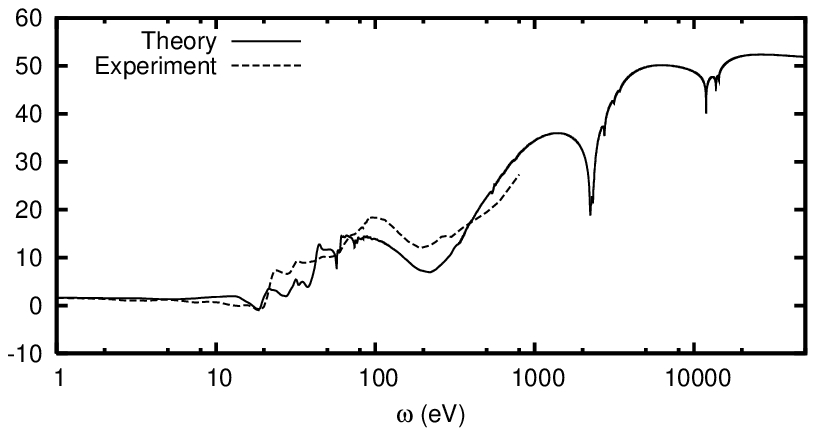}
 \caption
 {Calculated real part of the anomalous atomic scattering factor for diamond (top), Cu (middle), and Au
 (bottom) compared to experiment.~\cite{desy,hoc,cufp}} 
 \label{fprime}
\end{figure}

\begin{figure}
 \includegraphics{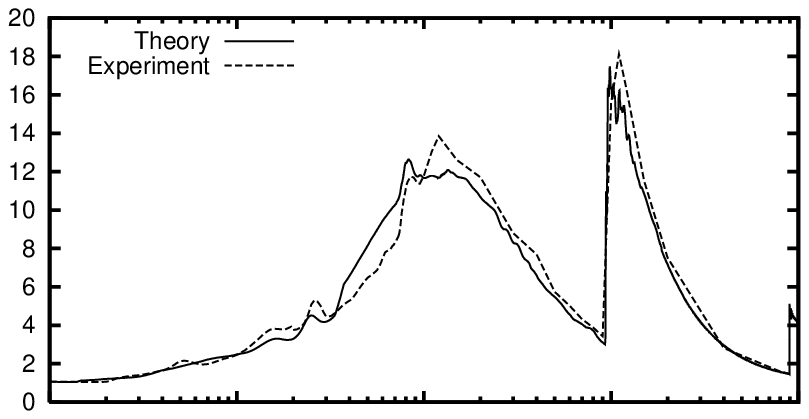}
 \includegraphics{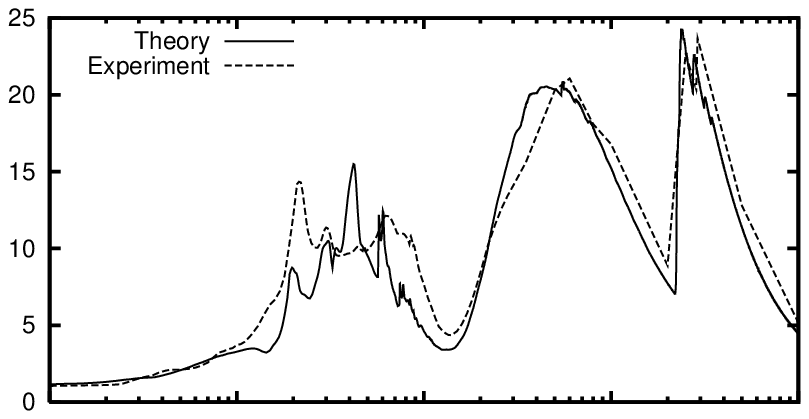}
 \includegraphics{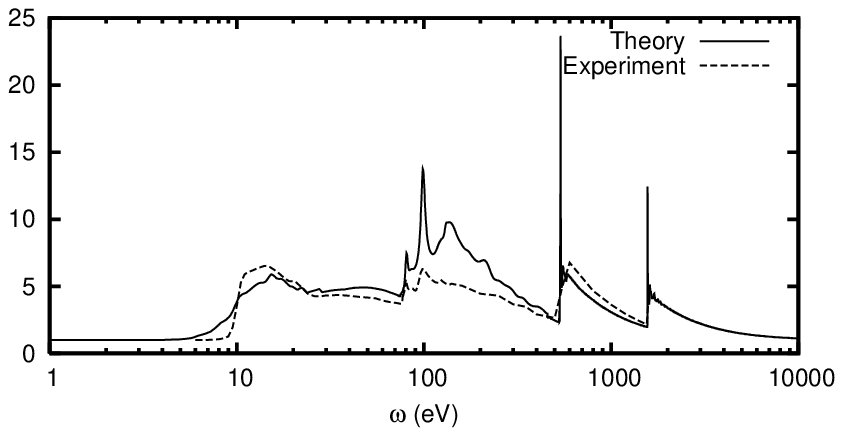}
 \caption
 {Calculated imaginary part of the atomic scattering factor for Cu (top),  Au
 (middle), and \alox (bottom) compared to experiment.~\cite{desy}} 
 \label{fdblprime_metal}
\end{figure}


\end{subsection}
\end{section}


\begin{section}{Applications and diagnositcs}

\begin{subsection}{Hamaker constant}
The Hamaker constant is the (real) function $\epsilon(i\omega)$ of a real
frequency $\omega$. For separation distances beyond the tunneling regime, the 
interaction between the tip and sample in an atomic force microscopy experiment
is dominated by the van der Waals force, which can be calculated given the
tip-sample geometry and the Hamaker constants of the tip and
sample.\cite{PhysRevB.42.1541} Using the analyticity of $\epsilon$ in the upper
half-plane, one can derive the following Kramers-Kronig type transform for
the Hamaker constant
\begin{equation}
  \epsilon(i\omega)=\frac{1}{\pi}\int_0^{\infty}d\omega\,' 
  \frac{\omega\,' \epsilon_2(\omega\,')+\omega \epsilon_1(\omega\,')}
  {\omega^2+\omega\,'^2}.
  \label{hamaker}
\end{equation}
We evaluate Eq.\ (\ref{hamaker}) numerically in the same way we evaluate
the Kramers-Kronig transform from $\epsilon_2$ to $\epsilon_1$, although away
from $\omega=0$ 
the integrand is regular.
  \end{subsection}

  \begin{subsection}{Sum rules}
\begin{figure}
 \includegraphics{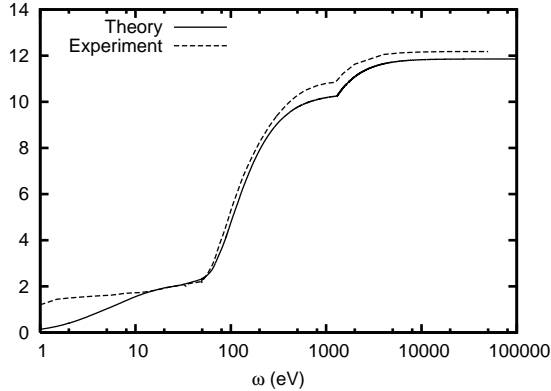}
 \caption
 {$n_{\rm eff}(\omega)$ calculated from the $\epsilon_2$ sum rule for Mg using
 Eq.\ (\ref{neff}), which assumes an asymptotic value of 12, the atomic number
 of Mg, in the $\omega \rightarrow \infty$ limit.} \label{sumMg}
 \end{figure}
Included in the output of our code are a few quantities useful for
understanding the relationship between the underlying electronic structure 
and the frequency dependance of the optical constants.
The $f$-sum rules for the imaginary parts of the dielectric function and the
inverse dielectric function provide an important quantitative check of the
calculation. We define the effective number of electrons per atom participating
in transitions at frequency $\omega$
\begin{equation}
  n_{\rm eff}(\omega)=\frac{V}{2 \pi^2 N} \int_0^{\omega} d\omega' \omega'
\epsilon_2 (\omega').
  \label{neff}
\end{equation}
This quantity has the limit~\cite{altarelli}
\begin{equation}
  \lim_{\omega \to\infty} n_{\rm eff}(\omega)=Z,
  \label{sumrule}
\end{equation}
where $Z$ is the number of electrons in the subsystem whose number density is
$N/V$. The theory and calculations presented here are valid over a frequency 
range large enough to quantitatively evaluate the limit (\ref{sumrule});
missing or extra oscillator strength implies invalid approximations or
unconverged calculations. Another check can be given by the index of
refraction sumrule.
\begin{equation}
  \int_0^{\infty} [n(\omega)-1]\, d\omega=0.
  \label{nsumrule}
\end{equation}
  \end{subsection}

\begin{subsection}{JDOS}
As stated above, the selection rules constrain the angular momentum of final
and initial states that can contribute to the absorption of light to a few 
channels (e.g. $p\rightarrow d$, $s\rightarrow p$, etc.). The joint density of states
(JDOS) correspondig to a certain dipole allowed chanel ($l \rightarrow l'$)
is defined in terms of the normal $l$-projected DOS $\rho_l$:
\begin{equation}
  \int_{E_f-\omega}^{E_f} \rho_{l}(E)\rho_{l'}(E+\omega)~dE,
\end{equation}
where the $l$-projected DOS is given in terms of the density matrix by
\begin{equation}
  \rho_l(E)=\sum_m \int d \vec r |Y_L(\hat r)|^2\rho (\vec r, \vec r, E).
\end{equation}
Neglecting energy dependence of the dipole matrix elements in the calculation
of $\epsilon_2$ gives a spectrum which is a sum of terms proportional to the
JDOS/$\omega^2$ for the dipole allowed channels. We show this quantity for
transitions from initial states with $p$ character compared to the calculated
$\epsilon_2$ for Diamond in Fig.~\ref{jdos}.

\begin{figure}
 \includegraphics{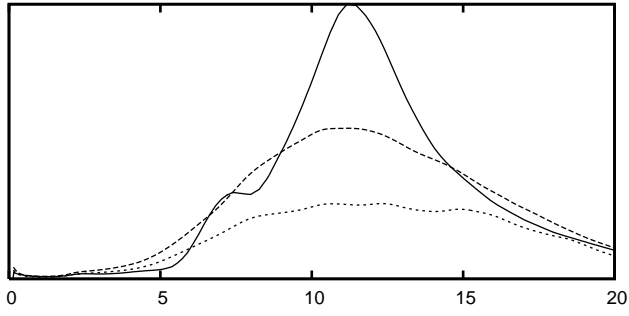}
 \caption
 {JDOS/$\omega^2$ for $p \rightarrow d$ (dashed line) and $p \rightarrow s$ (dots)
 transitions and the calculated $\epsilon_2$ of this work (solid line) for diamond vs. photon frequency in
 eV. } \label{jdos} 
 \end{figure}

\end{subsection}
\end{section}

\begin{section}{Conclusions}
  We have developed an efficient method for semi-quantitative
{\it ab initio} calculations of optical constants over a broad
spectrum, from the optical to x-ray frequencies.  Our method, based on the
one-particle density matrix, has been implemented in an extension of
the RSGF approach in the \feff\ codes which
can be applied to general, aperiodic materials. We have
illustrated the method here for a number of materials
for which optical data are also available including metals, insulators
and aperiodic solids.
Overall our results for the optical constants are semi-quantitative
in the optical-UV range, but become much more quantitative for
x-ray energies. Also their imaginary parts tend to be more accurate
compared to experiment.
 This degree of
accuracy is already adequate for many purposes, and especially for 
models which are not particularly sensitive to the detailed fine-structure
in the spectra such as the calculation of screened coulomb potentials
and van der Waals interactions.  Furthermore many improvements are possible:
i) It is desirable to include local field corrections
as described above;
ii) the muffin-tin approximation should  be replaced 
with more accurate {\it full potentials} in each cell;
iii) the extension to arbitrary momentum transfer $\vec q$ is
often desirable.
As noted above, the calculations can
  be done for any momentum transfer with only a modest increase in
  computational effort within our density-matrix formulation. In fact, the
  response of core states has already been extended to finite $q$ by
  Soininen, et. al; \cite{soininen}
iv) for crystalline systems, it may be desirable and sensible to calculate
the MS matrix in $k$-space, i.e., with periodic boundary conditions; and
v) the treatment of the particle-hole interaction $K$ currently 
only takes intra-atomic screening into account.
  
  \begin{acknowledgments}
We wish to thank A. Ankudinov, H. Lawler, G. Hug,
 E. Shirley, J. A. Soininen, A. Sorini, Y. Takimoto,
and F. Vila for many helpful discussions.
This work was supported in part by  DOE Grant DE-FG03-97ER45623 and
facilitated by the DOE CMSN.
  \end{acknowledgments}
\end{section}



\begin{thebibliography}{43}
\expandafter\ifx\csname natexlab\endcsname\relax\def\natexlab#1{#1}\fi
\expandafter\ifx\csname bibnamefont\endcsname\relax
  \def\bibnamefont#1{#1}\fi
\expandafter\ifx\csname bibfnamefont\endcsname\relax
  \def\bibfnamefont#1{#1}\fi
\expandafter\ifx\csname citenamefont\endcsname\relax
  \def\citenamefont#1{#1}\fi
\expandafter\ifx\csname url\endcsname\relax
  \def\url#1{\texttt{#1}}\fi
\expandafter\ifx\csname urlprefix\endcsname\relax\def\urlprefix{URL }\fi
\providecommand{\bibinfo}[2]{#2}
\providecommand{\eprint}[2][]{\url{#2}}

\bibitem[{\citenamefont{Onida et~al.}(2002)\citenamefont{Onida, Reining, and
  Rubio}}]{tddftvmbpt}
\bibinfo{author}{\bibfnamefont{G.}~\bibnamefont{Onida}},
  \bibinfo{author}{\bibfnamefont{L.}~\bibnamefont{Reining}}, \bibnamefont{and}
  \bibinfo{author}{\bibfnamefont{A.}~\bibnamefont{Rubio}},
  \bibinfo{journal}{Rev. Mod. Phys.} \textbf{\bibinfo{volume}{74}},
  \bibinfo{pages}{601} (\bibinfo{year}{2002}).

\bibitem[{\citenamefont{Nozi\`eres and Pines}(1958)}]{nozieres}
\bibinfo{author}{\bibfnamefont{P.}~\bibnamefont{Nozi\`eres}} \bibnamefont{and}
  \bibinfo{author}{\bibfnamefont{D.}~\bibnamefont{Pines}},
  \bibinfo{journal}{Phys. Rev.} \textbf{\bibinfo{volume}{109}},
  \bibinfo{pages}{741} (\bibinfo{year}{1958}).

\bibitem[{\citenamefont{Ehrenreich and Cohen}(1959)}]{ehrenreich}
\bibinfo{author}{\bibfnamefont{H.}~\bibnamefont{Ehrenreich}} \bibnamefont{and}
  \bibinfo{author}{\bibfnamefont{M.~H.} \bibnamefont{Cohen}},
  \bibinfo{journal}{Phys. Rev.} \textbf{\bibinfo{volume}{115}},
  \bibinfo{pages}{786} (\bibinfo{year}{1959}).

\bibitem[{\citenamefont{Adler}(1962)}]{adler}
\bibinfo{author}{\bibfnamefont{S.~L.} \bibnamefont{Adler}},
  \bibinfo{journal}{Phys. Rev.} \textbf{\bibinfo{volume}{126}},
  \bibinfo{pages}{413} (\bibinfo{year}{1962}).

\bibitem[{\citenamefont{Wiser}(1963)}]{wiser}
\bibinfo{author}{\bibfnamefont{N.}~\bibnamefont{Wiser}},
  \bibinfo{journal}{Phys. Rev.} \textbf{\bibinfo{volume}{129}},
  \bibinfo{pages}{62} (\bibinfo{year}{1963}).

\bibitem[{\citenamefont{Palik}(1985)}]{hoc}
\bibinfo{author}{\bibfnamefont{E.~D.} \bibnamefont{Palik}},
  \emph{\bibinfo{title}{Handbook of Optical Constants of Solids}}
  (\bibinfo{publisher}{Academic Press, Orlando}, \bibinfo{year}{1985}).

\bibitem[{\citenamefont{Elam et~al.}(2002)\citenamefont{Elam, Ravel, and
  Sieber}}]{elam}
\bibinfo{author}{\bibfnamefont{W.~T.} \bibnamefont{Elam}},
  \bibinfo{author}{\bibfnamefont{B.~D.} \bibnamefont{Ravel}}, \bibnamefont{and}
  \bibinfo{author}{\bibfnamefont{J.~R.} \bibnamefont{Sieber}},
  \bibinfo{journal}{Rad. Phys. Chem.} \textbf{\bibinfo{volume}{63}},
  \bibinfo{pages}{121} (\bibinfo{year}{2002}).

\bibitem[{\citenamefont{Henke et~al.}(1993)\citenamefont{Henke, Gullikson, and
  Davis}}]{henke}
\bibinfo{author}{\bibfnamefont{B.~L.} \bibnamefont{Henke}},
  \bibinfo{author}{\bibfnamefont{E.~M.} \bibnamefont{Gullikson}},
  \bibnamefont{and} \bibinfo{author}{\bibfnamefont{J.~C.} \bibnamefont{Davis}},
  \bibinfo{journal}{Atom. Data Nucl. Data Tables}
  \textbf{\bibinfo{volume}{54}}, \bibinfo{pages}{181} (\bibinfo{year}{1993}).

\bibitem[{\citenamefont{Yeh and Lindau}(1985)}]{yeh}
\bibinfo{author}{\bibfnamefont{J.~J.} \bibnamefont{Yeh}} \bibnamefont{and}
  \bibinfo{author}{\bibfnamefont{I.}~\bibnamefont{Lindau}},
  \bibinfo{journal}{Atomic Data and Nuclear Data Tables}
  \textbf{\bibinfo{volume}{32}}, \bibinfo{pages}{1} (\bibinfo{year}{1985}),
  \urlprefix\url{http://www.sciencedirect.com/science/article/B6WBB-4DBJ6HV-54%
/1/900634e5011c524427bd924083cdfc98}.

\bibitem[{\citenamefont{Zangwill and Soven}(1980)}]{zangsov}
\bibinfo{author}{\bibfnamefont{A.}~\bibnamefont{Zangwill}} \bibnamefont{and}
  \bibinfo{author}{\bibfnamefont{P.}~\bibnamefont{Soven}},
  \bibinfo{journal}{Phys. Rev. A} \textbf{\bibinfo{volume}{21}},
  \bibinfo{pages}{1561} (\bibinfo{year}{1980}).

\bibitem[{\citenamefont{Runge and Gross}(1984)}]{rungegross}
\bibinfo{author}{\bibfnamefont{E.}~\bibnamefont{Runge}} \bibnamefont{and}
  \bibinfo{author}{\bibfnamefont{E.~K.~U.} \bibnamefont{Gross}},
  \bibinfo{journal}{Phys. Rev. Lett.} \textbf{\bibinfo{volume}{52}},
  \bibinfo{pages}{997} (\bibinfo{year}{1984}).

\bibitem[{\citenamefont{Sham and Kohn}(1966)}]{shamkohn}
\bibinfo{author}{\bibfnamefont{L.~J.} \bibnamefont{Sham}} \bibnamefont{and}
  \bibinfo{author}{\bibfnamefont{W.}~\bibnamefont{Kohn}},
  \bibinfo{journal}{Phys. Rev.} \textbf{\bibinfo{volume}{145}},
  \bibinfo{pages}{561} (\bibinfo{year}{1966}).

\bibitem[{\citenamefont{Strinati}(1984)}]{strinati}
\bibinfo{author}{\bibfnamefont{G.}~\bibnamefont{Strinati}},
  \bibinfo{journal}{Phys. Rev. B} \textbf{\bibinfo{volume}{29}},
  \bibinfo{pages}{5718} (\bibinfo{year}{1984}).

\bibitem[{\citenamefont{Rohlfing and Louie}(2000)}]{rohlfing}
\bibinfo{author}{\bibfnamefont{M.}~\bibnamefont{Rohlfing}} \bibnamefont{and}
  \bibinfo{author}{\bibfnamefont{S.~G.} \bibnamefont{Louie}},
  \bibinfo{journal}{Phys. Rev. B} \textbf{\bibinfo{volume}{62}},
  \bibinfo{pages}{4927} (\bibinfo{year}{2000}).

\bibitem[{\citenamefont{Rivas}(2004)}]{gildardo}
\bibinfo{author}{\bibfnamefont{G.}~\bibnamefont{Rivas}}, Ph.D. thesis,
  \bibinfo{school}{University of {W}ashington} (\bibinfo{year}{2004}).

\bibitem[{\citenamefont{J.~D.~Jackson}(1975)}]{jackson}
\bibinfo{author}{\bibnamefont{J.~D.~Jackson}},
  \emph{\bibinfo{title}{Classical Electrodynamics}} (\bibinfo{publisher}{John
  Wiley \& Sons, Inc.}, \bibinfo{year}{1975}).

\bibitem[{\citenamefont{Reining et~al.}(2002)\citenamefont{Reining, Olevano,
  Rubio, and Onida}}]{PhysRevLett.88.066404}
\bibinfo{author}{\bibfnamefont{L.}~\bibnamefont{Reining}},
  \bibinfo{author}{\bibfnamefont{V.}~\bibnamefont{Olevano}},
  \bibinfo{author}{\bibfnamefont{A.}~\bibnamefont{Rubio}}, \bibnamefont{and}
  \bibinfo{author}{\bibfnamefont{G.}~\bibnamefont{Onida}},
  \bibinfo{journal}{Phys. Rev. Lett.} \textbf{\bibinfo{volume}{88}},
  \bibinfo{pages}{066404} (\bibinfo{year}{2002}).

\bibitem[{\citenamefont{Vasiliev et~al.}(1999)\citenamefont{Vasiliev,
  \"O\ifmmode~\breve{g}\else \u{g}\fi{}\"ut, and
  Chelikowsky}}]{PhysRevLett.82.1919}
\bibinfo{author}{\bibfnamefont{I.}~\bibnamefont{Vasiliev}},
  \bibinfo{author}{\bibfnamefont{S.}~\bibnamefont{\"O\ifmmode~\breve{g}\else
  \u{g}\fi{}\"ut}}, \bibnamefont{and} \bibinfo{author}{\bibfnamefont{J.~R.}
  \bibnamefont{Chelikowsky}}, \bibinfo{journal}{Phys. Rev. Lett.}
  \textbf{\bibinfo{volume}{82}}, \bibinfo{pages}{1919} (\bibinfo{year}{1999}).

\bibitem[{\citenamefont{Poiarkova and Rehr}(1999)}]{poiarkova}
\bibinfo{author}{\bibfnamefont{A.~V.} \bibnamefont{Poiarkova}}
  \bibnamefont{and} \bibinfo{author}{\bibfnamefont{J.~J.} \bibnamefont{Rehr}},
  \bibinfo{journal}{Phys. Rev. B} \textbf{\bibinfo{volume}{59}},
  \bibinfo{pages}{948} (\bibinfo{year}{1999}).

\bibitem[{\citenamefont{Grant}(1970)}]{grant}
\bibinfo{author}{\bibfnamefont{I.~P.} \bibnamefont{Grant}},
  \bibinfo{journal}{Advan. Phys.} \textbf{\bibinfo{volume}{19}},
  \bibinfo{pages}{747} (\bibinfo{year}{1970}).

\bibitem[{\citenamefont{Rehr and Albers}(2000)}]{review}
\bibinfo{author}{\bibfnamefont{J.~J.} \bibnamefont{Rehr}} \bibnamefont{and}
  \bibinfo{author}{\bibfnamefont{R.~C.} \bibnamefont{Albers}},
  \bibinfo{journal}{Rev. Mod. Phys.} \textbf{\bibinfo{volume}{72}},
  \bibinfo{pages}{621} (\bibinfo{year}{2000}).

\bibitem[{\citenamefont{Hedin and Lundqvist}(1969)}]{hl}
\bibinfo{author}{\bibfnamefont{L.}~\bibnamefont{Hedin}} \bibnamefont{and}
  \bibinfo{author}{\bibfnamefont{S.}~\bibnamefont{Lundqvist}},
  \emph{\bibinfo{title}{Solid state physics: advances in research and
  applications}} (\bibinfo{publisher}{Academic press, Inc.},
  \bibinfo{year}{1969}), vol.~\bibinfo{volume}{23}, pp.
  \bibinfo{pages}{1--181}.

\bibitem[{\citenamefont{Ankudinov and Rehr}(1997)}]{ankrer}
\bibinfo{author}{\bibfnamefont{A.~L.}~\bibnamefont{Ankudinov}} \bibnamefont{and}
  \bibinfo{author}{\bibfnamefont{J.~J.} \bibnamefont{Rehr}},
  \bibinfo{journal}{Phys. Rev. B} \textbf{\bibinfo{volume}{56}},
  \bibinfo{pages}{{R}1712} (\bibinfo{year}{1997}).

\bibitem[{\citenamefont{Ankudinov and Rehr}(2000)}]{scatfac}
\bibinfo{author}{\bibfnamefont{A.~L.}~\bibnamefont{Ankudinov}} \bibnamefont{and}
  \bibinfo{author}{\bibfnamefont{J.~J.} \bibnamefont{Rehr}},
  \bibinfo{journal}{Phys. Rev. B} \textbf{\bibinfo{volume}{62}},
  \bibinfo{pages}{002437} (\bibinfo{year}{2000}).

\bibitem[{\citenamefont{Tamura}(1992)}]{tamura}
\bibinfo{author}{\bibfnamefont{E.}~\bibnamefont{Tamura}},
  \bibinfo{journal}{Phys. Rev. B} \textbf{\bibinfo{volume}{45}},
  \bibinfo{pages}{3271} (\bibinfo{year}{1992}).

\bibitem[{\citenamefont{Ankudinov et~al.}(1998)\citenamefont{Ankudinov, Ravel,
  Rehr, and Conradson}}]{conrad}
\bibinfo{author}{\bibfnamefont{A.~L.} \bibnamefont{Ankudinov}},
  \bibinfo{author}{\bibfnamefont{B.}~\bibnamefont{Ravel}},
  \bibinfo{author}{\bibfnamefont{J.~J.} \bibnamefont{Rehr}}, \bibnamefont{and}
  \bibinfo{author}{\bibfnamefont{S.~D.} \bibnamefont{Conradson}},
  \bibinfo{journal}{Phys. Rev. B} \textbf{\bibinfo{volume}{58}},
  \bibinfo{pages}{7565} (\bibinfo{year}{1998}).

\bibitem[{\citenamefont{Ankudinov et~al.}(1996)\citenamefont{Ankudinov,
  Zabinsky, and Rehr}}]{ankzabrer}
\bibinfo{author}{\bibfnamefont{A.~L.} \bibnamefont{Ankudinov}},
  \bibinfo{author}{\bibfnamefont{S.~I.} \bibnamefont{Zabinsky}},
  \bibnamefont{and} \bibinfo{author}{\bibfnamefont{J.~J.} \bibnamefont{Rehr}},
  \bibinfo{journal}{Comp. Phys. Comm.} \textbf{\bibinfo{volume}{98}},
  \bibinfo{pages}{359} (\bibinfo{year}{1996}).

\bibitem[{\citenamefont{Zabinski et~al.}(1995)\citenamefont{Zabinski, Rehr,
  Ankudinov, Albers, and Eller}}]{zabin}
\bibinfo{author}{\bibfnamefont{S.~I.} \bibnamefont{Zabinsky}},
  \bibinfo{author}{\bibfnamefont{J.~J.} \bibnamefont{Rehr}},
  \bibinfo{author}{\bibfnamefont{A.}~\bibnamefont{Ankudinov}},
  \bibinfo{author}{\bibfnamefont{R.~C.} \bibnamefont{Albers}},
  \bibnamefont{and} \bibinfo{author}{\bibfnamefont{M.~J.} \bibnamefont{Eller}},
  \bibinfo{journal}{Phys. Rev. B} \textbf{\bibinfo{volume}{52}},
  \bibinfo{pages}{2995} (\bibinfo{year}{1995}).

\bibitem[{\citenamefont{Hageman et~al.}(1974)\citenamefont{Hageman, Gudat, and
  Kunz}}]{desy}
\bibinfo{author}{\bibfnamefont{H.}~\bibnamefont{Hageman}},
  \bibinfo{author}{\bibfnamefont{W.}~\bibnamefont{Gudat}}, \bibnamefont{and}
  \bibinfo{author}{\bibfnamefont{C.}~\bibnamefont{Kunz}}, \bibinfo{type}{Tech.
  Rep.}, \bibinfo{institution}{DESY} (\bibinfo{year}{1974}).

\bibitem[{\citenamefont{Newville}()}]{datacu}
\bibinfo{author}{\bibfnamefont{M.}~\bibnamefont{Newville}},
  \bibinfo{howpublished}{Private comunication}.

\bibitem[{\citenamefont{Waidmann et~al.}(2001)\citenamefont{Waidmann, Knupfer,
  Fink, Kleinsorge, and Robertson}}]{waidmann}
\bibinfo{author}{\bibfnamefont{S.}~\bibnamefont{Waidmann}},
  \bibinfo{author}{\bibfnamefont{M.}~\bibnamefont{Knupfer}},
  \bibinfo{author}{\bibfnamefont{J.}~\bibnamefont{Fink}},
  \bibinfo{author}{\bibfnamefont{B.}~\bibnamefont{Kleinsorge}},
  \bibnamefont{and}
  \bibinfo{author}{\bibfnamefont{J.}~\bibnamefont{Robertson}},
  \bibinfo{journal}{Journal of Applied Physics} \textbf{\bibinfo{volume}{89}},
  \bibinfo{pages}{3783} (\bibinfo{year}{2001}),
  \urlprefix\url{http://link.aip.org/link/?JAP/89/3783/1}.

\bibitem[{\citenamefont{Galli et~al.}(1989)\citenamefont{Galli, Martin, Car,
  and Parrinello}}]{galli}
\bibinfo{author}{\bibfnamefont{G.}~\bibnamefont{Galli}},
  \bibinfo{author}{\bibfnamefont{R.~M.} \bibnamefont{Martin}},
  \bibinfo{author}{\bibfnamefont{R.}~\bibnamefont{Car}}, \bibnamefont{and}
  \bibinfo{author}{\bibfnamefont{M.}~\bibnamefont{Parrinello}},
  \bibinfo{journal}{Phys. Rev. Lett.} \textbf{\bibinfo{volume}{62}},
  \bibinfo{pages}{555} (\bibinfo{year}{1989}).

\bibitem[{\citenamefont{Kresse and Furthm\"uller}(1996)}]{vasp}
\bibinfo{author}{\bibfnamefont{G.}~\bibnamefont{Kresse}} \bibnamefont{and}
  \bibinfo{author}{\bibfnamefont{J.}~\bibnamefont{Furthm\"uller}},
  \bibinfo{journal}{Phys. Rev. B} \textbf{\bibinfo{volume}{54}},
  \bibinfo{pages}{11169} (\bibinfo{year}{1996}).

\bibitem[{\citenamefont{Prange et~al.}(2007)\citenamefont{Prange, Rehr, and
  Rivas}}]{web_opcons}
\bibinfo{author}{\bibfnamefont{M.~P.} \bibnamefont{Prange}},
  \bibinfo{author}{\bibfnamefont{J.~J.} \bibnamefont{Rehr}}, \bibnamefont{and}
  \bibinfo{author}{\bibfnamefont{G.}~\bibnamefont{Rivas}},
  \emph{\bibinfo{title}{Full spectrum optical constants}}
  (\bibinfo{year}{2007}),
  \urlprefix\url{http://leonardo.phys.washington.edu/feff/opcons/}.

\bibitem[{\citenamefont{Soininen et~al.}(2005)\citenamefont{Soininen,
  Ankudinov, and Rehr}}]{soininen}
\bibinfo{author}{\bibfnamefont{J.~A.} \bibnamefont{Soininen}},
  \bibinfo{author}{\bibfnamefont{A.~L.} \bibnamefont{Ankudinov}},
  \bibnamefont{and} \bibinfo{author}{\bibfnamefont{J.~J.} \bibnamefont{Rehr}},
  \bibinfo{journal}{Physical Review B (Condensed Matter and Materials Physics)}
  \textbf{\bibinfo{volume}{72}}, \bibinfo{eid}{045136}
  (pages~\bibinfo{numpages}{10}) (\bibinfo{year}{2005}),
  \urlprefix\url{http://link.aps.org/abstract/PRB/v72/e045136}.

\bibitem[{\citenamefont{Lawler et~al.}(2008)\citenamefont{Lawler, Rehr, Vila,
  Dalosto, Shirley, and Levine}}]{lawler}
\bibinfo{author}{\bibfnamefont{H.~M.} \bibnamefont{Lawler}},
  \bibinfo{author}{\bibfnamefont{J.~J.} \bibnamefont{Rehr}},
  \bibinfo{author}{\bibfnamefont{F.}~\bibnamefont{Vila}},
  \bibinfo{author}{\bibfnamefont{S.~D.} \bibnamefont{Dalosto}},
  \bibinfo{author}{\bibfnamefont{E.~L.} \bibnamefont{Shirley}},
  \bibnamefont{and} \bibinfo{author}{\bibfnamefont{Z.~H.}
  \bibnamefont{Levine}}, \emph{\bibinfo{title}{Optical to uv spectra and
  birefringence of sio$_2$ and tio$_2$: First-principles calculations with
  excitonic effects}} (\bibinfo{year}{2008}),
  \urlprefix\url{http://www.citebase.org/abstract?id=oai:arXiv.org:0807.1920}.

\bibitem[{\citenamefont{Stratton}(1941)}]{stratton}
\bibinfo{author}{\bibfnamefont{J.}~\bibnamefont{Stratton}},
  \emph{\bibinfo{title}{Electromagnetic Theory}}
  (\bibinfo{publisher}{{McGraw-Hill}}, \bibinfo{year}{1941}).

\bibitem[{\citenamefont{Ankudinov and Rehr}(2000)}]{scatfact}
\bibinfo{author}{\bibfnamefont{A.~L.} \bibnamefont{Ankudinov}}
  \bibnamefont{and} \bibinfo{author}{\bibfnamefont{J.~J.} \bibnamefont{Rehr}},
  \bibinfo{journal}{Phys. Rev. B} \textbf{\bibinfo{volume}{62}},
  \bibinfo{pages}{2437} (\bibinfo{year}{2000}).

\bibitem[{\citenamefont{Cross et~al.}(1998)\citenamefont{Cross, Newville, Rehr,
  Sorensen, Bouldin, Watson, Gouder, Lander, and Bell}}]{cufp}
\bibinfo{author}{\bibfnamefont{J.~O.} \bibnamefont{Cross}},
  \bibinfo{author}{\bibfnamefont{M.}~\bibnamefont{Newville}},
  \bibinfo{author}{\bibfnamefont{J.~J.} \bibnamefont{Rehr}},
  \bibinfo{author}{\bibfnamefont{L.~B.} \bibnamefont{Sorensen}},
  \bibinfo{author}{\bibfnamefont{C.~E.} \bibnamefont{Bouldin}},
  \bibinfo{author}{\bibfnamefont{G.}~\bibnamefont{Watson}},
  \bibinfo{author}{\bibfnamefont{T.}~\bibnamefont{Gouder}},
  \bibinfo{author}{\bibfnamefont{G.~H.} \bibnamefont{Lander}},
  \bibnamefont{and} \bibinfo{author}{\bibfnamefont{M.~I.} \bibnamefont{Bell}},
  \bibinfo{journal}{Phys. Rev. B} \textbf{\bibinfo{volume}{58}},
  \bibinfo{pages}{11215} (\bibinfo{year}{1998}).

\bibitem[{\citenamefont{Hartmann}(1990)}]{PhysRevB.42.1541}
\bibinfo{author}{\bibfnamefont{U.}~\bibnamefont{Hartmann}},
  \bibinfo{journal}{Phys. Rev. B} \textbf{\bibinfo{volume}{42}},
  \bibinfo{pages}{1541} (\bibinfo{year}{1990}).

\bibitem[{\citenamefont{Altarelli et~al.}(1972)\citenamefont{Altarelli, Dexter,
  Nussenzveig, and Smith}}]{altarelli}
\bibinfo{author}{\bibfnamefont{M.}~\bibnamefont{Altarelli}},
  \bibinfo{author}{\bibfnamefont{D.~L.} \bibnamefont{Dexter}},
  \bibinfo{author}{\bibfnamefont{H.~M.} \bibnamefont{Nussenzveig}},
  \bibnamefont{and} \bibinfo{author}{\bibfnamefont{D.~Y.} \bibnamefont{Smith}},
  \bibinfo{journal}{Phys. Rev. B} \textbf{\bibinfo{volume}{6}},
  \bibinfo{pages}{4502} (\bibinfo{year}{1972}).

\bibitem[{\citenamefont{Rehr and Albers}(1990)}]{rehralb}
\bibinfo{author}{\bibfnamefont{J.~J.} \bibnamefont{Rehr}} \bibnamefont{and}
  \bibinfo{author}{\bibfnamefont{R.~C.} \bibnamefont{Albers}},
  \bibinfo{journal}{Phys. Rev. B} \textbf{\bibinfo{volume}{41}},
  \bibinfo{pages}{8139} (\bibinfo{year}{1990}).

\bibitem[{\citenamefont{Faulkner and Stocks}(1980)}]{faulknerstocks}
\bibinfo{author}{\bibfnamefont{J.~S.} \bibnamefont{Faulkner}} \bibnamefont{and}
  \bibinfo{author}{\bibfnamefont{G.~M.} \bibnamefont{Stocks}},
  \bibinfo{journal}{Phys. Rev. B} \textbf{\bibinfo{volume}{21}},
  \bibinfo{pages}{3222} (\bibinfo{year}{1980}).

\end{thebibliography}

\appendix*
\setcounter{equation}{0}
\section*{Appendix: Real-Space Multiple Scattering Green's Function}

In this Appendix we describe the real-space Green's functions used in this
work. Formally the Greens functions operator is given by
\begin{equation}
 G^+  (E) = [ E-H + i\delta]^{-1},
  \label{Gdef}
\end{equation}
where $\delta$ is
a positive infinitesimal.  Expanding $G^ +$ in the scattering potentials and free
propagators $G^{0}$ yields the multiple scattering (MS) expansion 
\begin{eqnarray}
  G= G^0+ G^0 VG =& G^0+G^0TG^0+\cdots \nonumber \\
                 =& [1-\bar G^0 T]^{-1} G^0 
  \label{dysonexp}
\end{eqnarray}
Here we have introduced the local $t$-matrix $t_n=v_n+v_nG^{0}t$ to sum
implicitly over all scatterings at a given site $n$,
where $\left < \vecr | t_n |
\vecrp \right > = t_n(\vecr,\vecrp,E)$ vanishes outside a given cell $n$ where
$v(r_n)$=0.

\begin{subsection}{Free propagator}

In position space the free propagator $G^0(E)$ is given by the FT,
\begin{equation}
  G^{0}(\vecr,\vecrp,E) = \int \frac{d^3 k}{ (2\pi)^3 }
  \frac {e^{i\vec k\cdot (\vecr - \vecrp)} }{E-\frac{k^2}{2} + i\delta}.
\end{equation}
Below we evaluate this expression in terms of site-angular momentum
scattering states $| L,R \rangle$  which diagonalize $t_i$
  \begin{equation}
    \begin{split}
j_L(\vecr_R) =& \langle \vecr  | L,R \rangle = i^l j_l(kr_R)Y_L(\hat r_R) \\
\bar j_L(\vecr_R) =& \langle L,R | \vecr \rangle = i^{-l} j_l(kr_R)Y_L^*(\hat r_R),
    \end{split}
  \end{equation}
  where $k=\sqrt{2(E-V_{0})}$.

In terms of spherical Bessel functions the free propagator is given
everywhere by
\begin{eqnarray}
  G^{0 }(\vecr,\vecrp,E) &=& -2k \sum_L Y_L(\hat r) g_l(r,r')
Y_L^*(\hat r') \\
&=& -2k \sum_L h^ + _L(\vecr_>) \bar j_L(\vecr_<),
\label{g0bess}
\end{eqnarray}
where $g_l(r,r')= h^+_l(kr_>) j_l(kr_<)$ and
$h^+_L(\vecr) = i^l h^+_l(kr) Y_L(\hat r)$.
This result can be obtained, e.g., from the FT using the identity 
$\exp(i\vec k\cdot\vecr)= 4\pi \Sigma_L j_L(\vecr) Y_L^*(\hat k)$
and carrying out the radial integrals in the complex $k$-plane.
Alternatively the same result follows from the inhomogeneous radial
differential equation, where
the prefactor is obtained from the Wronskian $2/r^2 W(j_l,h^+_l) = -2k$.
Here, as in the treatment of Rehr and Albers,~\cite{rehralb}
we have used the phase and normalization conventions of Messiah,
with $j_l=(h^+_l-h^-_l)/2i$ and $i^l h_l(x) = e^{ ix}c_l(1/ix)/x,$
$c_l$ is a polynomial of degree $l$ with $c_l(0)=1$.
Also, for convenience, we have included the phase factors
$i^l$ and $i^{-l}$ in $h^+_L$ and $\bar j_L$ respectively, which
do not change $G^0$, but simplify the asymptotic behavior.

The expansion of the free propagator for points at different
sites has the form of a matrix product
\begin{equation}
\begin{split}
  &G^{0 }(\vecr,\vecrp,E) =
  \sum_{L,L'} j_L(\vecr_R) G^{0 }_{LR,L'R'} \bar j_L(\vecr_{R'}) \\
&= 
\sum_{L,L'}\langle \vecr |LR\rangle\langle LR|G^{0 } (E)| L'R'\rangle
\langle L'R'|\vecrp\rangle.
\label{g0sep}
\end{split}
\end{equation}
This follows directly from Eq.\ (\ref{g0bess}) and the translation formulae for the
spherical Hankel functions \cite{rehralb}
\begin{equation}
  h^+ _{L'}(\vecrp_{R})= \sum_L j_L(\vecr_R)\, G^{0 }_{LR,L'R'}. 
\end{equation}
Note the implicit factors of $i^{l'}$ and $i^{l}$ in $j_L(\vecr_R)$
and $\bar j_{L'}(\vecr_R)$ in this representation.
In some works, e.g. that of Faulkner and Stocks \cite{faulknerstocks}, these 
phase factors are included in the definition the propagator matrix elements.
The above expression can be checked, e.g., by comparing
$i^l h^+_l(kr)= \sum_{L'} j_l'(kr_{R}) i^{l'} G^{0+}_{L'R,L0}$.
Eq.\ (\ref{g0sep}) can be derived, e.g., by expanding the exponential product
$ e^{i\vec k\cdot (\vecr - \vecrp)} =
 e^{i\vec k\cdot (\vecr -\vec R)}
 e^{-i\vec k\cdot (\vecrp -\vec R\,')}
 e^{i\vec k\cdot (\vec R -\vec R\,')}$ in spherical Bessel functions,
and then carrying out the integration over $k$. 
This procedure yields for the dimensionless propagator matrix elements:
\begin{equation}
  \tilde G^{0}_{LR,L'R'} \equiv  \frac{G^{0}_{LR,L'R'}}{-2k}
  = 4\pi \sum_{L''} \langle Y_L Y_{L''}| Y_{L'}\rangle\, h^ +_{L''}(k\vec R\,'').
\end{equation}
which depend explicitly on $k\vec R''= k(\vec R -\vec R\,')$. 
The FEFF code uses dimensionless matrix elements 
$\tilde G^0_{L,L'}(k\vec R)$ which have a separable
representation\cite{rehralb}
\begin{eqnarray}
&&\tilde G^0_{L,L'}(k\vec R) \equiv \tilde G^0_{LR,L'R'} =
\frac{e^{ikR}}{kR} \sum_{\lambda}
\tilde \Gamma_{L\lambda} \Gamma_{\lambda,L'}, \\
&&\rightarrow 4\pi \frac{e^{ikR}}{kR} c_l c_l' Y_L^*(\hat R) Y_{L'}(\hat R),
 \ (kR \rightarrow\infty) ,
\end{eqnarray}
where $\Gamma_{\lambda,L}(k \vec R)$ are generalized spherical harmonics.
This can be obtained, for example, by substituting the asymptotic form
of $i^lh_l$ and 
and the completeness relation
$\sum_{L}Y^*_{L}(\hat k) Y_{L}(\hat R) =\delta(\hat k -\hat R)$.

 \end{subsection}

\begin{subsection}{ Full propagator }

Let us now evaluate the behavior of the full propagator
$G(\vecr,\vecrp,E)$ for $\vecr$
and $\vecrp$ in different cells $n$ and $n'$ respectively.
For this case the MS series can be viewed as
a sequence of scattering events consisting of all scatterings
at site $n$ followed by all sequences of scatterings not scattering at site $n$
first or site $n'$ last, followed by all scatterings at site $n'$, 
\begin{equation}
  G_{nn'} =[1+G^{0}t_n]\bar G_{nn'}[1 + t_{n'} G^0],
  \label{alex210offdiag}
\end{equation}
where the notation $G_{nn'}$ refers to the propagator
starting and ending in cells $n$ and
$n'$ respectively, while $\bar G_{nn'}$ refers to those terms in the MS
expansion with first scatterings at sites other than $n$ and last scatterings
at sites other than $n'$.
This can be evaluated by substituting the representation of Eq.\ (\ref{g0sep})
into Eq.\ (\ref{alex210offdiag}) and then re-expressing the terms in the
site-angular momentum basis.
Then $\bar G_{nn'}$ can
be expressed in terms of the dimensionless full multiple scattering matrix
elements  $\bar G_{Ln,L'n'}$ where
  \begin{equation}
  \begin{split}
    \bar G(\vecr,\vecrp,E)
    &= \sum_{L,L'}j_L(\vecr_n)\, \bar G_{Ln,L'n'}\, \bar j_L(\vecr_{n'}) \\
    \bar G_{Ln,L'n'} &= \left[ 1-\bar G^{0} T \right] ^{-1}
    \bar G^{0} \big|_{Ln,L'n'}\,
    \label{FMS}
  \end{split}
  \end{equation}
  where $\bar G^{0}_{Ln,L'n'}=G^{0}_{Ln,L'n'} (1-\delta_{nn'}). $
The complementary delta-function in $\bar G^0$ ensures that $\bar G$
only includes initial scatterings from sites other than
$n$ and and final scatterings from sites other than $n'$. Next
the terms on the left and the right sides of Eq.\ (\ref{alex210offdiag}) can be expressed
in terms
of scattering states $R_{Ln}(\vecr_n)$.
To see this note that matrix elements of the 
dimensionless $t$-matrices can be expressed in terms of phase shifts as
\begin{equation}
  \begin{split}
    & \langle j_L| \tilde t_n| j_{L'} \rangle =\tilde t_{ln} \delta_{L,L'} \\
    & \tilde t_{ln} = e^{i\delta_{ln}}\sin\delta_{ln} .
  \end{split}
\end{equation}
Then using the representation of $G^0$ in
terms of Bessel functions in Eq.\ (\ref{g0bess}), one obtains
\begin{equation}
\begin{split}
  & \langle \vecr| [1+\tilde G^{0} \tilde t_n] |LR\rangle
  \equiv R_{Ln}(\vecr_n) e^{i\delta_{ln}} \\
&\ \ = i^l [j_l(r_n) +
h^+_l(r_n) \tilde t_{ln}] Y_L(\hat r_n), (r_n> r^{\rm mt}_n),
\end{split}
\end{equation}
where $R_L(\vecr) = i^l R_{ln}(r)Y_L(\hat r)$. Asymptotically 
$R_{ln}(r) = [ h^+_l e^{i\delta_{ln}} - h^-_l e^{-i\delta_{ln}} ]/2i \rightarrow
\sin (kr -l\pi/2 + \delta_{ln})/kr$.
For $ r <  r^{\rm mt}$, the radial states can be obtained from
the regular solution to the radial equation, matched to the above result.
Similarly one obtains
$\langle LR|(1+t_n G^{0})| \vecr \rangle = \bar R_L(\vecr_{n})\exp( i\delta_{ln})$.
Note that the radial functions $R_{ln}(r)$ in the scattering states
are real for real, nonnegative $k$, but are otherwise the
analytic continuation to complex $k$.
Combining all these results in Eq.\ (\ref{alex210offdiag}) then yields 
\begin{eqnarray}
  &&G(\vecr,\vecrp,E) = -2k  
  \nonumber \\ &\times& 
  \sum_{LL'} R_{Ln}(\vecr_n) 
 \tilde G_{Ln,L'n'} 
 \bar R_{L'n'}(\vecr_{n'});
  \nonumber \\  
 &&\tilde G_{Ln,L'n'} =
  e^{i\delta_{ln}} 
 \bar G_{Ln,L'n'} 
 e^{i\delta_{l'n'}}.
 \label{NSD G}
\end{eqnarray}
It is straightforward to show that this expression is equivalent to
that of Faulkner and Stocks \cite{faulknerstocks}.

For $\vecr$ and $\vecrp$ at the same site $n$, $G= G^{0} + G^{0}t_nG^{0} + 
\bar G_{n,n}$,  where $\bar G$ is given by Eq.\ (\ref{FMS}). This yields
\begin{eqnarray}
  && G(\vecr,\vecrp,E) = -2k \Big [
  \sum_L H_{Ln}(\vecr_>) \bar R_L(\vecr_<) \nonumber \\
\label{Gc}
  && + 
  \sum_{L,L'} R_{Ln}(\vecr_n) 
\tilde G_{Ln,L'n} 
\bar R_{L'n}(\vecr_{n}) \Big ],
 \label{site diagonal G}
\end{eqnarray}
where $H_L(\vecr)$ is the outgoing scattering
state at site $R$ which matches to
$ i^l e^{i\delta_{ln}} h^+_l(kr_n)$ for  $r_n >
r^{\rm mt}_n$.

\end{subsection}
\end{document}